\RequirePackage[2020-02-02]{latexrelease}
\documentclass{jfp}
\usepackage{proof}
\usepackage{amssymb}
\usepackage{txfonts}
\newtheorem{lemma}{Lemma}[section]

\newtheorem{definition}{Definition}[section]

\def\mod{\mbox{mod}}
\def\MRL{\mbox{MRL}}

\def\MRLJ{\mbox{MRLJ}}

\def\LMRL{\mbox{LMRL}}
\def\LMRLJ{\mbox{LMRLJ}}

\def\prim{a}
\def\fneg{f}
\def\limp{\supset}
\def\subst#1#2#3{#3[#2/#1]}
\def%
\mrl#1#2#3%
{#2\,{\land_#1}\,#3}
\def%
\mrlq#1#2%
{{\forall_#1}(#2)}
\def\interp#1#2{[#2]_{#1}}

\def\setcomp#1{{\overline#1}}
\def\jimp{\Rightarrow}
\def\ilimp{\Rightarrow}
\def\ilequ{\Leftrightarrow}
\def\fullset{\setcomp{\emptyset}}
\def\mpcut{\mbox{\it{mp-cut}}}
\def\mpcutconj{\mbox{\it{mp-cut-conj}}}
\def\mpcutdisj{\mbox{\it{mp-cut-disj}}}
\def\tpjg{\vdash}
\def\temd{\models}
\def\sizeof#1{|#1|}
\def\height#1{\mbox{\it ht}(#1)}
\def\formset#1{\{#1\}}

\title
{Multirole Logic and Multiparty Channels}

\author
[Hongwei Xi and Hanwen Wu]
{
{Hongwei Xi and Hanwen Wu} \\
Boston University, Boston, MA 02215, USA\\
\email{hwxi@cs.bu.edu, hwwu@cs.bu.edu}
}

\jdate{July, 2018}

\begin{document}
\maketitle

\begin%
{abstract}
We identify multirole logic as a new form of logic in which
conjunction/disjunction is interpreted as an ultrafilter on some
underlying set of roles and the notion of negation is generalized to
endomorphisms on this set.  We formulate both multirole logic (MRL)
and linear multirole logic (LMRL) as natural generalizations of
classical logic (CL) and classical linear logic (CLL),
respectively. Among various meta-properties established for MRL and
LMRL, we obtain one named multiparty cut-elimination stating that
every cut involving one or more sequents (as a generalization of a
binary cut involving exactly two sequents) can be eliminated, thus
extending the celebrated result of cut-elimination by Gentzen. As a
side note, we also give an ultrafilter-based interpretation for
intuitionism, formulating MRLJ as a natural generalization of
intuitionistic logic (IL). An immediate application of LMRL can be
found in a formulation of session types for channels that support
multiparty communication in distributed programming. We present a
multi-threaded lambda-calculus (MTLC) where threads communicate on
linearly typed multiparty channels that are directly rooted in LMRL,
establishing for MTLC both type preservation and global progress.  The
primary contribution of the paper consists of both identifying
multirole logic as a new form of logic and establishing a theoretical
foundation for it, and the secondary contribution lies in applying
multirole logic to the practical domain of distributed programming.
\end{abstract}

\section
{Introduction}
\label{sec:introduction}    
While the first and foremost inspiration for multirole logic
originates from studies on multiparty session types in distributed
programming, it seems natural in retrospective to introduce multirole
logic by exploring the well-known duality between conjunction and
disjunction in classical logic. For instance, in a two-sided
presentation of the classical sequent calculus (LK), we have the
following rules for conjunction and disjunction:
\def\Ggamma{\underline{A}}
\def\Ddelta{\underline{B}}
$$
\begin%
{array}{c}
\infer%
[\hbox{\bf(conj-r)\hss}]
{\Ggamma\tpjg\Ddelta, A\land B}
{\Ggamma\tpjg\Ddelta, A & \Ggamma\tpjg\Ddelta, B}
\\[6pt]
\infer%
[\hbox{\bf(conj-l-1)\hss}]
{\Ggamma, A\land B\tpjg\Ddelta}{\Ggamma, A\tpjg\Ddelta}
\kern18pt
\infer%
[\hbox{\bf(conj-l-2)\hss}]
{\Ggamma, A\land B\tpjg\Ddelta}{\Ggamma, B\tpjg\Ddelta}
\\[6pt]
\infer%
[\hbox{\bf(disj-l)\hss}]
{\Ggamma, A\lor B\tpjg\Ddelta}
{\Ggamma, A\tpjg\Ddelta & \Ggamma, B\tpjg\Ddelta}
\\[6pt]
\infer%
[\hbox{\bf(disj-r-1)\hss}]
{\Ggamma\tpjg\Ddelta, A\lor B}{\Ggamma\tpjg\Ddelta, A}
\kern18pt
\infer%
[\hbox{\bf(disj-r-2)\hss}]
{\Ggamma\tpjg\Ddelta, A\lor B}{\Ggamma\tpjg\Ddelta, B}
\\
\end{array}
$$
where $\Ggamma$ and $\Ddelta$ range over sequents (that are
essentially multisets of formulas).  One possible explanation of this
duality is to think of the availability of two roles $0$ and $1$ such
that the left side of a sequent judgment (of the form
$\Ggamma\tpjg\Ddelta$) plays role $1$ while the right side does role
$0$. In addition, there are two logical connectives $\land_0$ and
$\land_1$; $\land_r$ is given a conjunction-like interpretation by the
side playing role $r$ and disjunction-like interpretation by the other
side playing role $1-r$, where $r$ ranges over $0$ and $1$. With this
explanation, it seems entirely natural for us to introduce more roles
into classical logic.

\def\RC{\mbox{\underline{\rm R}}}
Multirole logic is parameterized over a chosen underlying set of
roles, which may be infinite, and we use $\fullset$ to refer to this
set. We use $R$ to range over role sets, which are just subsets of
$\fullset$.  Given any role set $R$, we use $\setcomp{R}$ for the
complement of $R$ in $\fullset$. Also, we use $R_1\uplus R_2$ for the
disjoint union of $R_1$ and $R_2$ (where $R_1\cap R_2=\emptyset$ is
assumed).

For the moment, let us assume that $\fullset$ consists all of the
natural numbers less than $N$ for some given $N\geq 2$.  Intuitively,
a conjunctive multirole logic is one in which there is a logical
connective $\land_r$ for each $r\in\fullset$ such that $\land_r$ is
given a conjunction-like interpretation by a side playing role $r$ and
a disjunction-like interpretation otherwise.  If we think of the
universal quantifier $\forall$ as an infinite form of conjunction,
then what is said about $\land$ can be readily applied to $\forall$ as
well. In fact, additive, multiplicative, and exponential connectives
in linear logic~\cite{LinearLogic} can all be treated in a similar
manner.  Dually, a disjunctive multirole logic can be formulated (by
giving $\land_r$ a disjunction-like interpretation if the side plays
the role $r$ and a conjunction-like interpretation otherwise). We
primarily focus on conjunctive multirole logic in this paper.

Given a formula $A$ and a set $R$ of roles, we write $\interp{R}{A}$
for an i-formula, which is some sort of interpretation of $A$ based on
$R$. For instance, the interpretation of $\land_r$ based on $R$ is
conjunction-like if $r\in R$ holds, and it is disjunction-like
otherwise.  A crucial point, which we take from studies on multiparty
session types, is that interpretation should be based on sets of
roles rather than just individual roles. In other words, one side is
allowed to play multiple roles simultaneously. A sequent $\Gamma$ in
multirole logic is a multiset of i-formulas, and such a sequent is
inherently many-sided as each $R$ appearing in $\Gamma$ represents
precisely one side. As can be readily expected, the (binary) cut-rule
in (either conjunctive or disjunctive) multirole logic is of the
following form:
$$
\begin%
{array}{l}
\infer%
{\Gamma_1,\Gamma_2}
{\Gamma_1,\interp{R}{A} & \Gamma_2,\interp{\setcomp{R}}{A}}
\end{array}
$$
The cut-rule can be interpreted as some sort of communication between
two parties in distributed
programming~\cite{Abramsky94,BellinS94,CairesPfenning10,Wadler12-Prop-as-Sessions}. For
communication between multiple parties, it seems natural to seek a
generalization of the cut-rule that involves more than two sequents.
In conjunctive multirole logic, the admissibility of the following
rule, which is given the name $\mpcutconj(n)$, can be established:
$$
\begin%
{array}{l}
\infer%
{\tpjg\Gamma_1,\ldots,\Gamma_n}
{\setcomp{R_1}\uplus\cdots\uplus\setcomp{R_n}=\fullset &
 \tpjg\Gamma_1,\interp{R_1}{A} & \cdots & \tpjg\Gamma_n,\interp{R_n}{A}
}
\end{array}
$$
In disjunctive multirole logic, the admissibility of the following
rule, which is given the name $\mpcutdisj(n)$, can be established:
$$
\begin%
{array}{l}
\infer%
{\tpjg\Gamma_1,\ldots,\Gamma_n}
{R_1\uplus\cdots\uplus R_n=\fullset &
 \tpjg\Gamma_1,\interp{R_1}{A} & \cdots & \tpjg\Gamma_n,\interp{R_n}{A}
}
\end{array}
$$
We may use the name $\mpcut$ to refer to either $\mpcutconj$ or
$\mpcutdisj$, which itself is a shorthand for $\mbox{multiparty-cut}$.

\def\lneg{\lnot}
In classical logic, the negation operator is clearly one of a
kind. With respect to negation, the conjunction and disjunction
operators behave dually, and the universal and existential quantifiers
behave dually as well.  For the moment, let us write $\lneg{A}$ for
the negation of $A$.  It seems rather natural to interpret
$\interp{R}{\lneg{A}}$ as $\interp{\setcomp{R}}{A}$. Unfortunately,
such an interpretation of negation immediately breaks $\mpcut(n)$ for
any $n\geq 3$ (and it breaks $\mpcut(1)$ as well). What we discover
regarding negation is a bit of surprise: The notion of negation can be
generalized to endomorphisms on the underlying set $\fullset$ of
roles. For instance, if $f$ maps $i$ to $(i+1)~\mod~{3}$ for
$i=0,1,2$, then the negation operator $\lneg_{f}$ based on $f$ is of
order $3$, that is, $A$ and $\lneg_{f}(\lneg_{f}(\lneg_{f}(A)))$ are
equivalent for any formula $A$ in $\MRL$ (which is a multirole version
of CL)

Furthermore, we incorporate the notion of multirole into
intuitionistic logic as well as linear logic, formulating $\MRLJ$ and
$\LMRL$ as multirole versions of IL and CLL, respectively.  An
immediate application of multirole logic can be found in the practical
domain of distributed programming, where we make direct use of $\LMRL$
in a design of linearly typed channels for supporting communication
between multiple parties.

The rest of the paper is organized as follows. In
Section~\ref{section:MRL}, we formulate $\MRL$, a multirole version of
first-order classical predicate logic, and then mention some key
meta-properties of $\MRL$. In particular, we formally state the
admissibility of a cut-rule (mp-cut) involving $n$ sequents for any
$n\geq 1$. In Section~\ref{section:MRLJ}, we present an
ultrafilter-based interpretation for intuitionism, formulating $\MRLJ$
as a multirole version of first-order intuitionistic predicate
logic. We move onto formulating $\LMRL$ in Section~\ref{section:LMRL}
as a multirole version of first-order linear classical predicate
logic. In Section~\ref{section:LMRL_session_types}, we point out a
profound relation between $\LMRL$ and session types for multiparty
channels. Subsequently, we present in Section~\ref{section:MTLCZ} and
Section~\ref{section:MTLCO} a multi-threaded lambda-calculus equipped
with linearly typed channels for multiparty communication, where the
types for channels are directly rooted in $\LMRL$, and briefly mention
some implementation work in Section~\ref{section:implementation}.
Lastly, we compare with some closely related work and then conclude.

\section
{MRL: Multirole Logic}
\label{section:MRL}
Let $\fullset$ be the underlying (possibly infinite) set of roles for
the multirole logic MRL presented in this section. Strictly speaking,
this MRL should be referred to as {\em first-order classical multirole
  predicate logic\/}.

\def\F{\mathcal{F}}
\def\U{\mathcal{U}}
\def\J{\mathcal{J}}
\def%
\invfun#1{\overline{#1}}
\def%
\prefun#1{{#1}^{-{\kern-1pt}1}}
\def\preimg#1#2{\prefun{#1}(#2)}
\def%
\mrln#1#2%
{{\lneg_#1}(#2)}
\def%
\mrlc#1#2#3%
{#2\,{\land_#1}\,#3}
\def%
\mrlq#1#2%
{{\forall_#1}(#2)}
\def%
\mrlimp#1#2#3#4%
{#3\,{\limp_{#1,#2}{#4}}}

\begin%
{definition}
A filter $\F$ on $\fullset$ is a subset of the power set of $\fullset$
such that
\begin%
{itemize}
\item $\fullset\in\F$
\item $R_1\in\F$ and $R_1\subseteq R_2$ implies $R_2\in\F$
\item $R_1\in\F$ and $R_2\in\F$ implies $R_1\cap R_2\in\F$
\end{itemize}
\end{definition}
A filter $\F$ on $\fullset$ is an ultrafilter if either $R\in\F$ or
$\setcomp{R}\in\F$ holds for every subset $R$ of $\fullset$.  We use
$\U$ to range over ultrafilters on $\fullset$.  When there is no risk
of confusion, we may simply use $r$ for the principal ultrafilter at
$r$, which is defined as $\{R\subseteq\fullset\mid r\in R\}$. If
$\fullset$ is finite, then it can be readily proven that each $\U$ on
$\fullset$ is a principal filter at some element $r\in\fullset$.

\def\myset#1{\{#1\}}
\def\power{\mathscr{P}}
\begin%
{definition}

Given an endomorphism $f$ on $\fullset$, that is, a mapping from
$\fullset$ to itself, we use $f(R)$ for the set $\myset{f(r)\mid r\in
  R}$ (image of $R$ under $f$) and $\prefun{f}(R)$ for the set
$\myset{r\mid f(r)\in R}$ (pre-image of $R$ under $f$). Clearly,
$\prefun{f}(\emptyset)=\emptyset$ and $\prefun{f}(\fullset)=\fullset$,
and $\prefun{f}$ is distributive over the following operations on
sets: union, intersection, and complement.
\end{definition}
\begin%
{proposition}
Given $f$ and $\F$, we use $f(\F)$ for the set
$\myset{R\mid\prefun{f}(R)\in\F}$.  It is clear that
$f(\F)$ is a filter (as $\F$ is).  If $\F$ is an ultrafilter, then
$f(\F)$ is also an ultrafilter.
\end{proposition}

Given an endomorphism $f$ on $\fullset$, we use $\lneg_{\fneg}$ for a
unary connective.  Given an ultrafilter $\U$ on $\fullset$, we use
$\land_{\U}$ for a binary connective and $\forall_{\U}$ for a
quantifier. Note that we may equally choose the name $\lor_{\U}$ for
$\land_{\U}$ (and $\exists_{\U}$ for $\forall_{\U}$) as the meaning of
the named connective (quantifier) solely comes from the ultrafilter
$\U$.

We use $t$ for first-order terms, which are standard (and thus not
formulated explicitly for brevity).  The formulas in $\MRL$ are
defined as follows:
\[
\begin%
{array}
{lrcl}
\mbox{formulas} & A,B & ::= & %
\prim \mid \mrln{\fneg}{A} \mid \mrlc{\U}{A}{B} \mid \mrlq{\U}{\lambda x.A} \\
\end{array}
\]
where $\prim$ ranges over pre-defined primitive formulas.
Instead of writing something like $\forall_{\U} x.A$, we write
$\forall_{\U}(\lambda x.A)$, where $x$ is a bound variable. Given a
formula $A$, a term $t$ and a variable $x$, we use $\subst{x}{t}{A}$
for the result of substituting $t$ for $x$ in $A$ and treat it as a
proper subformula of $\forall_{\U}(\lambda x.A)$.  For notational
convenience, we may also write $f(A)$ for $\mrln{\fneg}{A}$, $\U(A_1,
A_2)$ for $\mrlc{\U}{A_1}{A_2}$, and $\U(\lambda x.A)$ for
$\mrlq{\U}{\lambda x.A}$.

Note that there is no connective for implication in $\MRL$.  If
desired, one may simply treat $\mrlimp{\fneg}{\U}{A}{B}$ as
$\U(\fneg(A), B)$. We are only to introduce $\limp_{\fneg,\U}$ as
a primitive connective corresponding to implication for each given
pair of $\fneg$ and $\U$ when formulating an intuitionistic version of
$\MRL$.

\begin{figure}
\[
\begin%
{array}{c}
\infer%
[\hbox to 0pt{\bf(Id)\hss}]
{\tpjg\interp{R_1}{\prim},\ldots,\interp{R_n}{\prim}}
{\fullset=R_1\uplus\ldots\uplus R_n}
\\[6pt]
\infer%
[\hbox to 0pt{\bf(Weaken)\hss}]
{\tpjg\Gamma,\interp{R}{A}}{\tpjg\Gamma}
\\[6pt]
\infer%
[\hbox to 0pt{\bf(Contract)\hss}]
{\tpjg\Gamma,\interp{R}{A}}
{\tpjg\Gamma,\interp{R}{A},\interp{R}{A}}
\\[6pt]
\infer%
[\hbox to 0pt{\bf($\lneg$)\hss}]
{\tpjg\Gamma,\interp{R}{\mrln{\fneg}{A}}}
{\tpjg\Gamma,\interp{\preimg{\fneg}{R}}{A}}
\\[6pt]
\infer%
[\hbox to 0pt{\bf($\land$-neg-l)\hss}]
{\tpjg\Gamma,\interp{R}{\mrlc{\U}{A}{B}}}
{R\not\in\U & \tpjg\Gamma,\interp{R}{A}}
\\[6pt]
\infer%
[\hbox to 0pt{\bf($\land$-neg-r)\hss}]
{\tpjg\Gamma,\interp{R}{\mrlc{\U}{A}{B}}}
{R\not\in\U & \tpjg\Gamma,\interp{R}{B}}
\\[6pt]
\infer%
[\hbox to 0pt{\bf($\land$-pos)\hss}]
{\tpjg\Gamma,\interp{R}{\mrlc{\U}{A}{B}}}
{
R\in\U &
\tpjg\Gamma,\interp{R}{A} & \tpjg\Gamma,\interp{R}{B}
}
\\[6pt]
\infer%
[\hbox to 0pt{\bf($\forall$-neg)\hss}]
{\tpjg\Gamma,\interp{R}{\mrlq{\U}{\lambda x.A}}}
{R\not\in\U & \tpjg\Gamma,\interp{R}{\subst{x}{t}{A}}}
\\[6pt]
\infer%
[\hbox to 0pt{\bf($\forall$-pos)\hss}]
{\tpjg\Gamma,\interp{R}{\mrlq{\U}{\lambda x.A}}}
{R\in\U & x\not\in\Gamma & \tpjg\Gamma,\interp{R}{A}}
\\[6pt]
\end{array}
\]
\caption%
{The inference rules for $\MRL$}\label{fig:MRL:infrules}
\end{figure}
Given a formula $A$ and a set $R$ of roles, $\interp{R}{A}$ is
referred to as an i-formula (for interpretation of $A$ based on
$R$). Let us use $\Gamma$ for multisets of i-formulas, which are also
referred to as sequents.  The inference rules for $\MRL$ are listed
in Figure~\ref{fig:MRL:infrules}. In the rule
$\mbox{\bf($\forall$-pos)}$, $x\not\in\Gamma$ means that $x$ does not
have any free occurrences in i-formulas contained inside $\Gamma$.
Please note that a sequent $\Gamma$ in this formulation is many-sided
(rather than one-sided) as every $R$ appearing $\Gamma$ represents
precisely one side.

\def\D{\mathcal{D}}
Let us use $\sizeof{A}$ for the size of $A$, which is the number of
connectives contained in $A$, and use $\D$ for derivations of
sequents, which are just trees containing nodes that are applications
of inference rules.  Given a derivation $\D$, $\height{\D}$ stands for
the tree height of $\D$.  When writing $\D::\Gamma$, we mean that $\D$
is a derivation of $\Gamma$, that is, $\Gamma$ is the conclusion of
$\D$. We may use the following format to present an inference rule:
$$(\Gamma_1;\ldots;\Gamma_n)\jimp\Gamma_0$$ where $\Gamma_i$ for
$1\leq i\leq n$ are the premisses of the rule and $\Gamma_0$ the
conclusion. Such an inference rule is said to be admissible if its
conclusion is derivable whenever its premisses are.  Given two
i-formulas $\interp{R_1}{A}$ and $\interp{R_2}{B}$, we write
$\interp{R_1}{A}\ilimp\interp{R_2}{B}$ to mean that the rule
$(\Gamma,\interp{R_1}{A})\jimp(\Gamma,\interp{R_2}{B})$ is
admissible. And we write $A\ilimp B$ if
$\interp{R}{A}\ilimp\interp{R}{B}$ holds for any role set $R$. For
instance, for any ultrafilter $\U$ and formulas $A$ and $B$, we have
$\U(A, B)\ilimp\U(B, A)$.  In addition, we write
$\interp{R_1}{A}\ilequ\interp{R_2}{B}$ for both
$\interp{R_1}{A}\ilimp\interp{R_2}{B}$ and
$\interp{R_2}{B}\ilimp\interp{R_1}{A}$, and $A\ilequ B$ for both
$A\ilimp B$ and $B\ilimp A$.

As a new form of logic,
MRL may not seem intuitive. We present as follows some
simple properties to facilitate the understanding of MRL.
\begin%
{proposition}
\label{prop:neg-injective}
For each injective endomorphism $f$ on $\fullset$,
we have $\interp{R}{A}\ilimp\interp{f(R)}{f(A)}$ for any formula $A$ and
role set $R$.
\end{proposition}
\begin%
{proof}
By the rule~\mbox{\bf($\lneg$)}, we have
$\interp{\prefun{f}(f(R))}{A}\ilimp\interp{f(R)}{f(A)}$.
Since $f$ is injective, we have $R=\prefun{f}(f(R))$.
\end{proof}

Given an endomorphism $f$ on $\fullset$, we refer to $f$ as a
permutation if it is actually a bijection and use $\invfun{f}$ for the
inverse permutation of $f$. Please note that
$\invfun{f}(R)=\prefun{f}(R)$ for any role set $R$.

\begin%
{proposition}
\label{prop:neg-permutation}
Given any A, we have $A\ilimp\invfun{f}(f(A))$ for each permutation
$f$ on $\fullset$.
\end{proposition}
\begin%
{proof}
This proposition follows from
Proposition~\ref{prop:neg-injective} as $\invfun{f}(f(R))=R$ for any role set $R$.
\end{proof}

\def\id{{\it id}}
Given two endomorphisms $f_1$ and $f_2$ on $\fullset$, we write
$f_1\cdot f_2$ for a composition of $f_1$ and $f_2$ such that
$(f_1\cdot f_2)(r)=f_2(f_1(r))$ for $r\in\fullset$.  The set of
permutations on $\fullset$ forms a group where the group unit is the
identity function $\id$, and the group inverse of $f$ is $\invfun{f}$,
and the group product of $f_1$ and $f_2$ is $f_1\cdot f_2$. Given a
natural number $n$, $f^{n}$ is defined to be $\id$ if $n=0$ and
$f\cdot{f^{n-1}}$ if $n > 0$.  The order of $f$, if exists, is the
least positive integer $n$ such that $f^{n}=\id$.

\begin%
{proposition}
Assume that $\fullset$ is finite.
For each permutation $f$ on $\fullset$, we have $A\ilequ
f^{n}(A)$ for some $n\geq 1$.
\end{proposition}
\begin%
{proof}
By a theorem of Lagrange, every permutation $f$ on a finite $\fullset$
is of some order $n$ (such that $n$ divides $N!$ for $N$ being the
cardinality of $\fullset$).  By Proposition~\ref{prop:neg-injective},
we have $\interp{R}{A}\ilimp\interp{f^{n}(R)}{f^{n}(A)}$.  Since
$f^{n}(R)=R$, we have $\interp{R}{A}\ilimp\interp{R}{f^{n}(A)}$ (for
any $R$).  In other words, we have $A\ilimp f^{n}(A)$.  For any $R$,
we have $\tpjg\interp{R}{A},\interp{\setcomp{R}}{A}$ (by
Lemma~\ref{lemma:MRL:axiom}); we can derive
$\tpjg\interp{R}{A},\interp{\setcomp{R}}{f^{n}(A)}$ by
Proposition~\ref{prop:neg-injective}. Assume that $\tpjg\Gamma,
\interp{R}{f^{n}(A)}$ is derivable. By cut-elimination
(Lemma~\ref{lemma:MRL:mp-cut}), we can derive
$\tpjg\Gamma,\interp{R}{A}$. Hence, we have $f^{n}(A)\ilimp A$.
\end{proof}

\def\LK{\mbox{LK}}
The classical sequent calculus $\LK$ of Gentzen is a special case of MRL
where $\fullset=\myset{0,1}$. In a two-sided formulation of $\LK$, let
us assume that the left side plays role $1$ and the right side does
role $0$; the negation operator $\lneg$ is $\lneg_{\fneg}$ for
$\fneg=(0\;1)$, that is, $\fneg(0)=1$ and $\fneg(1)=0$; the
conjunction operator $\land$ is $\land_0$, where $0$ refers to the
principal ultrafilters at $0$, and the disjunction operator $\lor$ is
$\land_1$; the universal quantifier $\forall$ is $\forall_0$, and the
existential quantifier $\exists$ is $\forall_1$. Clearly, the order of the
permutation $(0\;1)$ is $2$. Therefore, we have $A\ilequ\lneg(\lneg(A))$.

\begin%
{proposition}
\label{prop:equivalences}
Given a permutation $f$ and an endomorphism $g$, we
write $f(g)$ for the endomorphism ${f}\cdot{g}\cdot\invfun{f}$. We have
the following equivalences:
$$\begin%
{array}{lrcl}
\mbox{(1)} & f(g(A)) & \Leftrightarrow & f(g)(f(A)) \\
\mbox{(2)} & f(\U(A, B)) & \Leftrightarrow & f(\U)(f(A), f(B)) \\
\mbox{(3)} & f(\U(\lambda x.A)) & \Leftrightarrow & f(\U)(\lambda x.f(A)) \\
\mbox{(4)} & \U_1(A, \U_2(B_1, B_2)) & \Leftrightarrow & \U_2(\U_1(A,B_1), \U_1(A,B_2)) \\
\end{array}$$
\end{proposition}
\begin%
{proof}
The (straightforward) proof details are omitted for brevity. Notice
that the two De~Morgan's laws in classical propositional logic are
unified in (2). Also, please notice that the fact is captured in (4)
that conjunction is distributive over disjunction and vice versa.
\end{proof}
Clearly, Proposition~\ref{prop:equivalences} suggests that many
notions of duality in mathematics can and should be revisited in the
context of multirole logic.

\subsection%
{Cut-Elimination for MRL}
We establish in this section that $\MRL$ enjoys a form of
cut-elimination possibly involving $n$ sequents for any $n\geq 1$.
Essentially, we are to prove the following lemma:

\begin
{lemma}
[mp-cut]
\label{lemma:MRL:mp-cut}
Let $R_1,\ldots,R_n$ be subsets of $\fullset$ for some $n\geq 1$.
If $\setcomp{R}_1\uplus\cdots\uplus\setcomp{R}_n=\fullset$ holds, then the following inference
rule $\mbox{(mp-cut)}$ is admissible in $\MRL$:
$$(\Gamma_1,\interp{R_1}{A};\ldots;\Gamma_n,\interp{R_n}{A})\jimp(\Gamma_1,\ldots,\Gamma_n)$$
\end{lemma}

Let us first present some crucial properties of $\MRL$ as follows:

\begin%
{lemma}
\label{lemma:MRL:fullset}
The following rule is admissible in $\MRL$:
$$
()\jimp{\interp{\fullset}{A}}
$$
\end{lemma}
\begin%
{proof}
By structural induction on $A$.
Note that we need the fact $\prefun{f}(\fullset)=\fullset$ in the
case where $A$ is of the form $\lneg_{f}(B)$.
\end{proof}

Let us use $\formset{\interp{R}{A}}$ for a sequent that contains an
indefinite number of occurrences of $\interp{R}{A}$.

\begin%
{lemma}
(Splitting of Roles)
\label{lemma:MRL:roleset-split}
The following rule is admissible in $\MRL$:
$$
(\Gamma,\interp{R_1\uplus R_2}{A})
\jimp
{\Gamma,\interp{R_1}{A},\interp{R_2}{A}}
$$
\end{lemma}
\begin%
{proof}
We can directly prove the admissibility of the following rule
by structural induction on $A$:
$$
(\Gamma
,\formset{\interp{R_1\uplus R_2}{A}})
\jimp
{\Gamma,\interp{R_1}{A},\interp{R_2}{A}}
$$
\end{proof}
Please note that the opposition of Lemma~\ref{lemma:MRL:roleset-split}
is not valid, that is, the rule
$(\Gamma,\interp{R_1}{A},\interp{R_2}{A})\jimp\Gamma,\interp{R_1\uplus
  R_2}{A}$ is not admissible.

\begin%
{lemma}
\label{lemma:MRL:axiom}
Assume
${R}_1\uplus\ldots\uplus{R}_n=\fullset$.
The following rule is admissible in $\MRL$:
$$
()\jimp{\interp{R_1}{A},\ldots,\interp{R_n}{A}}
$$
\end{lemma}
\begin%
{proof}
By Lemma~\ref{lemma:MRL:fullset} and Lemma~\ref{lemma:MRL:roleset-split}.
\end{proof}

\begin%
{lemma}
(1-cut)
\label{lemma:MRL:1-cut}
The following rule is admissible in $\MRL$:
$$
(\Gamma,\interp{\emptyset}{A})\jimp
{\Gamma}
$$
\end{lemma}
\begin%
{proof}
We can directly prove the admissibility of the following rule
by structural induction on $A$:
$$
(\Gamma,\formset{\interp{\emptyset}{A}})\jimp{\Gamma}
$$
\end{proof}
Note that Lemma~\ref{lemma:MRL:1-cut} can be seen as a special form
of cut-elimination where only one sequent is involved.

\begin%
{lemma}
\label{lemma:MRL:2-cut-residual}
(2-cut with residual)
Assume that $\setcomp{R_1}$ and $\setcomp{R_2}$
are disjoint.  Then the following rule (2-cut-residual) is admissible
in $\MRL$:
$$
(\Gamma_1,\interp{R_1}{A};
 \Gamma_2,\interp{R_2}{A})\jimp
{\Gamma_1,\Gamma_2,\interp{R_1\cap R_2}{A}}
$$
\end{lemma}
\begin%
{proof}
We omit the proof for this lemma as it can be done in essentially the
same manner as is the presented proof of
Lemma~\ref{lemma:LMRL:2-cut-residual} (which is simply the counterpart
of the current lemma expressed in the setting of linear multirole
logic).
\end{proof}

We present as follows a proof of Lemma~\ref{lemma:MRL:mp-cut} based
Lemma~\ref{lemma:MRL:1-cut} and
Lemma~\ref{lemma:MRL:2-cut-residual}:
\begin%
{proof}
The proof proceeds by induction on $n$.
If $n=1$,
then this lemma is just
Lemma~\ref{lemma:MRL:1-cut}.
Assume that $n\geq 2$ holds. Then we have
$\D_i::(\Gamma_i,\interp{R_i}{A})$ for $1\leq i\leq n$.
Clearly, $\setcomp{R_1}$ and $\setcomp{R_2}$ are disjoint.
By Lemma~\ref{lemma:MRL:2-cut-residual},
we have $\D_{12}::(\Gamma_1,\Gamma_2,\interp{{R_1\cap R_2}}{A})$.
By induction hypothesis, we can derive the sequent $\Gamma_1,\Gamma_2,\ldots,\Gamma_n$
based on $\D_{12},\ldots,\D_n$.
\end{proof}

This given proof of Lemma~\ref{lemma:MRL:mp-cut} clearly
indicates that multiparty cut-elimination builds on top of
Lemma~\ref{lemma:MRL:1-cut} (1-cut) and
Lemma~\ref{lemma:MRL:2-cut-residual} (2-cut-residual).  In
particular, one may see Lemma~\ref{lemma:MRL:1-cut} and
Lemma~\ref{lemma:MRL:2-cut-residual} as two fundamental
meta-properties of a logic.

It should be clear that the rule (2-cut) (that is, the special case of
(mp-cut) for $2$ sequents) plus Lemma~\ref{lemma:MRL:roleset-split}
implies the rule (2-cut-residual): Assume we have
$\tpjg\Gamma_1,\interp{R_1}{A}$ and $\tpjg\Gamma_2,\interp{R_2}{A}$
for some $R_1$ and $R_2$ such that $\setcomp{R}_1$ and $\setcomp{R}_2$
are disjoint; by applying Lemma~\ref{lemma:MRL:roleset-split} to
$\tpjg\Gamma_1,\interp{R_1}{A}$, we have
$\tpjg\Gamma_1,\interp{\setcomp{R_2}}{A},\interp{R_1\cap R_2}{A}$; by
applying (2-cut) to
$\tpjg\Gamma_1,\interp{\setcomp{R_2}}{A},\interp{R_1\cap R_2}{A}$ and
$\tpjg\Gamma_2,\interp{R_2}{A}$, we have
$\tpjg\Gamma_1,\Gamma_2,\interp{R_1\cap R_2}{A}$.

Also, it should be clear that the rule (3-cut) (that is, the special
case of (mp-cut) for $3$ sequents) immediately implies the rule
(2-cut-residual): Assume we have $\tpjg\Gamma_1,\interp{R_1}{A}$ and
$\tpjg\Gamma_2,\interp{R_2}{A}$ for some $R_1$ and $R_2$ such that
$\setcomp{R}_1$ and $\setcomp{R}_2$ are disjoint; we also have
$\tpjg\interp{R_3}{A},\interp{\setcomp{R_3}}{A}$ for $R_3=R_1\cap R_2$
by Lemma~\ref{lemma:MRL:axiom}; therefore we have
$\tpjg\Gamma_1,\Gamma_2,\interp{R_3}{A}$ by (3-cut) (as
$\setcomp{R}_1\uplus\setcomp{R}_2\uplus{R}_3=\fullset$). This is a
useful observation as what we actually implement in practice is the
rule (3-cut) (for $\LMRL$).

\section%
{MRLJ: Intuitionistic MRL}
\label{section:MRLJ}
If in classical logic the most prominent logical connective is
negation, then it is implication in intuitionistic logic. Given an
endomorphism $f$ and an ultrafilter $\U$ (on $\fullset$), we introduce
a logical connective $\limp_{f,\U}$ in $\MRLJ$ (for representing the
notion of implication). The formulas in $\MRLJ$ are
defined as follows:
\[
\begin%
{array}
{lrcl}
\mbox{formulas} & A,B & ::= & %
\prim \mid \mrln{\fneg}{A} \mid
\mrlc{\U}{A}{B} \mid \mrlimp{f}{\U}{A}{B}\mid \mrlq{\U}{\lambda x.A} \\
\end{array}
\]
We may also write $\U_{f}(A, B)$ for $\mrlimp{f}{\U}{A}{B}$.

\begin%
{definition}
\label{def:intuitionistic}  
Given an ultrafilter $\U$, a sequent $\Gamma$ is $\U$-intuitionistic
if there is at most one i-formula $\interp{R}{A}$ in $\Gamma$ such
that $R\in\U$ holds.  An inference rule $(\Gamma_1, \ldots,
\Gamma_n)\jimp\Gamma_0$ is $\U$-intuitionistic if $\Gamma_0$ and
$\Gamma_1,\ldots,\Gamma_n$ are $\U$-intuitionistic.
\end{definition}

Intuitionistic multirole logic is parameterized over a fixed
ultrafilter on $\fullset$. In $\MRLJ$, we refer to this ultrafilter as
$\J$.  Every inference rule in Figure~\ref{fig:MRL:infrules} is also
an inference rule in $\MRLJ$ if it is $\J$-intuitionistic. In addition,
we have the following rules for handling implication:

$$
\begin%
{array}{c}
\infer%
[\hbox to 0pt{\bf($\limp$-neg)\hss}]
{\tpjg\Gamma,\interp{R}{\mrlimp{\fneg}{\U}{A}{B}}}
{R\not\in\U & \tpjg\Gamma,\interp{\preimg{\fneg}{R}}{A},\interp{R}{B}}
\\[6pt]
\infer%
[\hbox to 0pt{\bf($\limp$-pos)\hss}]
{\tpjg\Gamma_1,\Gamma_2,\interp{R}{\mrlimp{\fneg}{\U}{A}{B}}}
{R\in\U & \tpjg\Gamma_1,\interp{\preimg{\fneg}{R}}{A} & \tpjg\Gamma_2,\interp{R}{B}}
\\[6pt]
\end{array}
$$

\def\LJ{\mbox{LJ}}
The intuitionistic sequent calculus $\LJ$ of Gentzen is a special case
of MRLJ where $\fullset=\myset{0,1}$ and $\J$ is the principal filter
at $1$; the implication connective $\limp$ is $\limp_{f,\U}$ for
$f=(0\; 1)$ and $\U$ being the principal filter at $0$. As an example,
the well-known fact that $A\limp\lneg{B}$ implies $B\limp\lneg{A}$ in
$\LJ$ can be generalized in $\MRLJ$ as follows:
\begin%
{proposition}
Given two permutations $f$ and $g$,
we have $\U_{f}(A, g(B))\ilequ\U_{g}(B, f(A))$.
\end{proposition}
\begin%
{proof}
It is a routine to verify that the following sequent is derivable in $\MRLJ$
for any role set $R$:
$$(\interp{\setcomp{R}}{\U_{f}(A, g(B))}, \interp{R}{\U_{g}(B, f(A))})$$
Assume that $\tpjg\Gamma,\interp{R}{\U_{f}(A, g(B))}$ is
derivable. We have $\tpjg\Gamma, \interp{R}{\U_{g}(B, f(A))}$ by
cut-elimination (Lemma~\ref{lemma:MRLJ:mp-cut}).  Hence, $\U_{f}(A,
g(B))\ilimp\U_{g}(B, f(A))$ holds.  By symmetry, we thus have obtained
$\U_{f}(A,g(B))\ilequ\U_{g}(B, f(A))$.
\end{proof}

In $\LJ$, $\interp{R}{A}\ilimp\interp{R}{B}$ can be internalized as
$\interp{R}{A\limp B}$ if $R=\myset{1}$. This form of internalization
can be generalized in $\MRLJ$ as follows:

\begin%
{proposition}
Assume that $\tpjg\interp{R}{\U_{f}(A, B)}$
is derivable for $R\in\J$ and $R\not\in\U$ and $\prefun{f}(R)=\setcomp{R}$.
Then $\interp{R}{A}\ilimp\interp{R}{B}$ holds.
\end{proposition}
\begin%
{proof}
It is a routine to verify that the following sequent is derivable in $\MRLJ$:
$$(\interp{\setcomp{R}}{\U_{f}(A, B)}, \interp{\setcomp{R}}{A}, \interp{R}{B}$$
Assume that $\tpjg\Gamma,\interp{R}{A}$ is derivable.
By cut-elimination for $\MRLJ$ (Lemma~\ref{lemma:MRLJ:mp-cut}), we can
derive $\tpjg\Gamma,\interp{R}{B}$. Hence, we have $\interp{R}{A}\ilimp\interp{R}{B}$.
\end{proof}
    
\begin
{lemma}
[mp-cut]
\label{lemma:MRLJ:mp-cut}
Let $R_1,\ldots,R_n$ be subsets of $\fullset$ for some $n\geq
1$.  If $\setcomp{R}_1\uplus\cdots\uplus\setcomp{R}_n=\fullset$ holds,
then the following inference rule $\mbox{(mp-cut)}$ is admissible in
$\MRLJ$:
$$(\Gamma_1,\interp{R_1}{A};\ldots;\Gamma_n,\interp{R_n}{A})\jimp(\Gamma_1,\ldots,\Gamma_n)$$
where it is assumed that each $(\Gamma_i,\interp{R_i}{A})$ is $\J$-intuitionistic
for $1\leq i\leq n$.
\end{lemma}
\begin%
{proof}
The lemma can be proven in essentially the same fashion as is
Lemma~\ref{lemma:MRL:mp-cut}.
\end{proof}

We see the very ability to incorporate intuitionism into multirole
logic as solid evidence in support of multirole being an inherent
notion in logic.
  
\section%
{LMRL: Linear Multirole Logic}
\label{section:LMRL}
\def\aconj{{\&}}
\def\adisj{{\oplus}}
\def\mconj{{\otimes}}
\def\mdisj{{\invamp}}
\def%
\lmrlac#1#2#3%
{#2\,{\aconj_{#1}{#3}}}
\def%
\lmrlmc#1#2#3%
{#2\,{\mconj_{#1}{#3}}}
\def\bang{{!}}
\def\qmark{{?}}
\def\lmrlx#1#2{\bang_{#1}({#2})}
\begin{figure}
\[
\begin%
{array}{c}
\infer%
[\hbox to 0pt{\bf(Id)\hss}]
{\tpjg\interp{R_1}{\prim},\ldots,\interp{R_n}{\prim}}
{\fullset=R_1\uplus\ldots\uplus R_n}
\\[6pt]
\infer%
[\hbox to 0pt{\bf($\lneg$)\hss}]
{\tpjg\Gamma,\interp{R}{\mrln{\fneg}{A}}}
{\tpjg\Gamma,\interp{\preimg{\fneg}{R}}{A}}
\\[6pt]
\infer%
[\hbox to 0pt{\bf($\aconj$-neg-l)\hss}]
{
\tpjg\Gamma,\interp{R}{\lmrlac{\U}{A}{B}}
}
{R\not\in\U & \tpjg\Gamma,\interp{R}{A}}
\\[6pt]
\infer%
[\hbox to 0pt{\bf($\aconj$-neg-r)\hss}]
{
\tpjg\Gamma,\interp{R}{\lmrlac{\U}{A}{B}}
}
{R\not\in\U & \tpjg\Gamma,\interp{R}{B}}
\\[6pt]
\infer%
[\hbox to 0pt{\bf($\aconj$-pos)\hss}]
{\tpjg\Gamma,\interp{R}{\lmrlac{\U}{A}{B}}}
{
R\in\U &
\tpjg\Gamma,\interp{R}{A} & \tpjg\Gamma,\interp{R}{B}
}
\\[6pt]
\infer%
[\hbox to 0pt{\bf($\mconj$-neg)\hss}]
{\tpjg\Gamma,\interp{R}{\lmrlmc{\U}{A}{B})}}
{R\not\in\U & \tpjg\Gamma,\interp{R}{A},\interp{R}{B}}
\\[6pt]
\infer%
[\hbox to 0pt{\bf($\mconj$-pos)\hss}]
{\tpjg\Gamma_1,\Gamma_2,\interp{R}{\lmrlmc{\U}{A}{B}}}
{
R\in\U &
\tpjg\Gamma_1,\interp{R}{A} & \tpjg\Gamma_2,\interp{R}{B}
}
\\[6pt]
\infer%
[\hbox to 0pt{\bf($\bang$-pos)\hss}]
{\tpjg\qmark(\Gamma)
,\interp{R}{\lmrlx{\U}{A}}}
{R\in\U & \tpjg\qmark(\Gamma),\interp{R}{A}}
\\[6pt]
\infer%
[\hbox to 0pt{\bf($\bang$-neg-weaken)\hss}]
{\tpjg\Gamma,\interp{R}{\lmrlx{\U}{A}}}
{R\not\in\U & \tpjg\Gamma}
\\[6pt]
\infer%
[\hbox to 0pt{\bf($\bang$-neg-derelict)\hss}]
{\tpjg\Gamma,\interp{R}{\lmrlx{\U}{A}}}
{R\not\in\U & \tpjg\Gamma,\interp{R}{A}}
\\[6pt]
\infer%
[\hbox to 0pt{\bf($\bang$-neg-contract)\hss}]
{\tpjg\Gamma,\interp{R}{\lmrlx{\U}{A}}}
{R\not\in\U & \tpjg\Gamma,\interp{R}{\lmrlx{\U}{A}},\interp{R}{\lmrlx{\U}{A}}}
\\[6pt]
\infer%
[\hbox to 0pt{\bf($\forall$-neg)\hss}]
{\tpjg\Gamma,\interp{R}{\mrlq{\U}{\lambda x.A}}}
{R\not\in\U & \tpjg\Gamma,\interp{R}{\subst{x}{t}{A}}}
\\[6pt]
\infer%
[\hbox to 0pt{\bf($\forall$-pos)\hss}]
{\tpjg\Gamma,\interp{R}{\mrlq{\U}{\lambda x.A}}}
{R\in\U & x\not\in\Gamma & \tpjg\Gamma,\interp{R}{A}}
\\[6pt]
\end{array}
\]
\caption%
{The inference rules for LMRL}\label{fig:LMRL:infrules}
\end{figure}
In this section, we generalize classical linear logic (CLL) to linear
multirole logic (LMRL), which is to guide subsequently a design of
linearly typed channels for multiparty communication in distributed
programming.

Let $\fullset$ be the full set of roles for $\LMRL$.  For each
endomorphism $f$ on $\fullset$, we assume a corresponding unary
connective $\lneg_{f}$ (generalized negation). For each ultrafilter
$\U$ on $\fullset$, we assume two binary connectives $\aconj_{\U}$ and
$\mconj_{\U}$ (corresponding to $\aconj$/$\adisj$ and
$\mconj$/$\mdisj$ in linear logic, respectively) and a unary
connective $\bang_{\U}$ (corresponding to $\bang$/$\qmark$ in linear
logic) and a quantifier $\forall_{\U}$ (corresponding to
$\forall$/$\exists$). The formulas in $\LMRL$ are defined as follows:
\[
\begin%
{array}
{lrcl}
\mbox{formulas} & A,B & ::= & %
\prim \mid \mrln{\fneg}{A} \mid \lmrlmc{\U}{A}{B} \mid \lmrlac{\U}{A}{B} \mid \lmrlx{\U}{A} \mid \mrlq{\U}{\lambda x.A} \\
\end{array}
\]
We may write
$\interp{R}{\qmark(A)}$ to mean $\interp{R}{\lmrlx{\U}{A}}$ for some
$R\not\in\U$, and $?(\Gamma)$ to mean that each i-formula in $\Gamma$
is of the form $\interp{R}{\qmark(A)}$.

The inference rules for $\LMRL$ are listed in
Figure~\ref{fig:LMRL:infrules}. Note that there are one rule for each
$\lneg_{f}$, and one positive rule and two negative rules for each
$\aconj_{\U}$, and one positive rule and one negative rule for each
$\mconj_{\U}$, and one positive rule and three negative rules for each
$\bang_{\U}$, and one positive rule and one negative rule for each
$\forall_{\U}$. Let us take the rule $\mbox{\bf($\aconj$-neg-l)}$ as
an example; the i-formula $\interp{R}{\lmrlac{\U}{A}{B}}$ is referred
to as the major i-formula of the rule.  Let us take the rule
$\mbox{\bf($\mconj$-pos)}$ as another example; the i-formula
$\interp{R}{\lmrlmc{\U}{A}{B}}$ is referred to as the major i-formula
of the rule. The major i-formulas for the other rules (excluding the
rule $\mbox{\bf(Id)}$) should be clear as well. For the rule
$\mbox{\bf(Id)}$, each $\interp{R_i}{a}$ is referred to as a major
i-formula.

Please recall that $\sizeof{A}$ is for the size of $A$ that is, the
number of connectives contained in $A$, and $\D$ for a derivation tree
and $\height{\D}$ for the height of the tree.  Also, we use
$\D::\Gamma$ for a derivation of $\Gamma$.

The following lemmas in $\LMRL$ are the counterparts of
of a series of lemmas
(Lemma~\ref{lemma:MRL:fullset},
Lemma~\ref{lemma:MRL:roleset-split}, Lemma~\ref{lemma:MRL:axiom},
Lemma~\ref{lemma:MRL:1-cut}, Lemma~\ref{lemma:MRL:2-cut-residual}, and
Lemma~\ref{lemma:MRL:mp-cut}) in $\MRL$.

\begin%
{lemma}
\label{lemma:LMRL:fullset}
The following rule is admissible in $\LMRL$:
$$
()\jimp{\interp{\fullset}{A}}
$$
\end{lemma}
\begin%
{proof}
By structural induction on $A$.  
\end{proof}

Given an i-formula $\interp{R}{A}$, let us use
$\formset{\interp{R}{A}}$ for a sequent consisting of only
$\interp{R}{A}$ if $A$ is not of the form $\lmrlx{\U}{B}$ for
$R\not\in \U$ or some repeated occurrences of $\interp{R}{A}$
otherwise (that is, if $A$ is of the form $\lmrlx{\U}{B}$ for
$R\not\in\U$).

\begin%
{lemma}
(Splitting of Roles)
\label{lemma:LMRL:roleset-split}
The following rule is admissible in $\LMRL$:
$$
(\Gamma,\interp{R_1\uplus R_2}{A})
\jimp
{\Gamma,\interp{R_1}{A},\interp{R_2}{A}}
$$
\end{lemma}
\begin%
{proof}
We can directly prove the admissibility of the following rule
by structural induction on $A$:
$$
(\Gamma
,\formset{\interp{R_1\uplus R_2}{A}})
\jimp
{\Gamma,\interp{R_1}{A},\interp{R_2}{A}}
$$
\end{proof}

\begin%
{lemma}
\label{lemma:LMRL:axiom}  
Assume
${R}_1\uplus\ldots\uplus{R}_n=\fullset$.
The following rule is admissible in $\LMRL$:
$$
()\jimp{\interp{R_1}{A},\ldots,\interp{R_n}{A}}
$$
\end{lemma}
\begin%
{proof}
By Lemma~\ref{lemma:LMRL:fullset} and Lemma~\ref{lemma:LMRL:roleset-split}.
\end{proof}

\begin%
{lemma}
(1-cut)
\label{lemma:LMRL:1-cut}
The following rule is admissible in $\LMRL$:
$$
(\Gamma,\interp{\emptyset}{A})\jimp
{\Gamma}
$$
\end{lemma}
\begin%
{proof}
We can directly prove the admissibility of the following rule
by structural induction on $A$:
$$
(\Gamma,\formset{\interp{\emptyset}{A}})\jimp{\Gamma}
$$
\end{proof}

\begin%
{lemma}
\label{lemma:LMRL:2-cut-residual}
(2-cut with residual)
Assume that
$\setcomp{R_1}$
and
$\setcomp{R_2}$
are disjoint.
Then the following rule is admissible in $\LMRL$:
$$
(\Gamma_1,\interp{R_1}{A};
 \Gamma_2,\interp{R_2}{A})\jimp
{\Gamma_1,\Gamma_2,\interp{R_1\cap R_2}{A}}
$$
\end{lemma}
\begin%
{proof}
The proof for this lemma is given in
Section~\ref{subsection:LMRL:2-cut-residual}.
\end{proof}

\begin
{lemma}
[mp-cut]
\label{lemma:LMRL:mp-cut}
Let $R_1,\ldots,R_n$ be subsets of $\fullset$ for some $n\geq 1$.
If $\setcomp{R}_1\uplus\cdots\uplus\setcomp{R}_n=\fullset$ holds, then the following inference
rule $\mbox{(mp-cut)}$ is admissible in $\LMRL$:
$$(\Gamma_1,\interp{R_1}{A};\ldots;\Gamma_n,\interp{R_n}{A})\jimp(\Gamma_1,\ldots,\Gamma_n)$$
\end{lemma}
\begin%
{proof}
By
Lemma~\ref{lemma:LMRL:1-cut} and
Lemma~\ref{lemma:LMRL:2-cut-residual},
the lemma can be proven in the same fashion as is Lemma~\ref{lemma:MRL:mp-cut}.
\end{proof}

\def\piLMRL{\mbox{$\pi$LMRL}}
By following some recent work on encoding cut-elimination as reduction
in variants of
$\pi$-calculus~\cite{CairesPfenning10,Wadler12-Prop-as-Sessions}, we
can readily formulate a variant of $\pi$-calculus in which process
reduction follows precisely the strategy employed in the proof of
Lemma~\ref{lemma:LMRL:2-cut-residual} for reducing the complexity of a
cut-formula. Alternatively, we can interpret formulas in $\LMRL$ as
session types for channels in a multi-threaded lambda-calculus and
various meta-properties of $\LMRL$ as certain (primitive) functions on
channels. We take the alternative here as it is closer to the goal
of implementing session-typed multiparty channels for practical use.

\subsection%
{Proof of Lemma~\ref{lemma:LMRL:2-cut-residual}}
\label{subsection:LMRL:2-cut-residual}
Due to the presence of the the structural rules
$\mbox{\bf($\bang$-neg-weaken)}$ and
$\mbox{\bf($\bang$-neg-contract)}$, we need to prove a
strengthened version of Lemma~\ref{lemma:LMRL:2-cut-residual} stating
that the following rule is admissible in $\LMRL$:
$$
(\Gamma_1,\formset{\interp{R_1}{A}};
 \Gamma_2,\formset{\interp{R_2}{A}})\jimp
{\Gamma_1,\Gamma_2,\interp{R_1\cap R_2}{A}}
$$
where we use $\formset{\interp{R}{A}}$ for a sequent consisting of
only $\interp{R}{A}$ if $A$ is not of the form $\lmrlx{\U}{B}$ for
$R\not\in\U$ or some repeated occurrences of $\interp{R}{A}$ if $A$ is
of the form $\lmrlx{\U}{B}$ for $R\not\in\U$.  The proof strategy we
use is essentially adopted from the one in a proof of cut-elimination
for classical linear logic (CLL)~\cite{LL-Troelstra92}.
Please see Section~\ref{Appendix:subsection:LMRL:2-cut-residual} in
Appendix for details.

\section%
{LMRL and Session Types}
\label{section:LMRL_session_types}
In broad terms, a session is a sequence of interactions between
two or more concurrently running processes and a session type is a
form of type for specifying or classifying the interactions in a
session.  In this section, we are to illustrate a profound relation
between LMRL and session types by mostly relying on (informal)
explanation and examples.  In particular, we are to outline the manner
in which logical connectives in LMRL and various type constructors for
session types are related.

\def\CH{\mbox{CH}}
We use $A$ and $B$ in the rest of this section both for formulas in
$\LMRL$ and for session types (which are formally defined in
Section~\ref{section:MTLCO}).  We use $\CH$ for (logical) channels
supporting communication between multiple parties (processes) involved
in a session. How such channels can be implemented based on some forms
of physical channels is beyond the scope of this paper. It is
entirely possible that distinct logical channels can share one
underlying physical channel.

\def\timp{\rightarrow}
\def\tchan{\mbox{\bf chan}}
\def\tunit{\mbox{$\mathbf 1$}}
\def\tnone{\mbox{\bf nil}}
\def\toption{\mbox{\bf option}}
\def\trepseq{\mbox{\bf repseq}}
\def\trepeat{\mbox{\bf repeat}}
We assume that each channel $\CH$ consists of one or more
endpoints and each process holding one endpoint can communicate
with the processes holding the other endpoints (of the same channel).
We use $\CH_{R}$ to denote one endpoint of $\CH$, where $R$ is a role
set (that is, a subset of $\fullset$).
\emph%
{At any given time, if $\CH$ consists of $n$ endpoints
$\CH_{R_1},\ldots,\CH_{R_n}$, then ${R}_1\uplus\cdots\uplus{R}_n=\fullset$}.
We say that a channel $\CH$ is of some session type $A$ if each
endpoint $\CH_{R}$ can be assigned the type $\tchan(R, A)$ (where
$\tchan$ is some linear type constructor).  We may write $\tchan(A)$
to mean $\tchan(R,A)$ for some role set $R$.  We see this view of a
channel and its endpoints as a style of interpretation for
Lemma~\ref{lemma:LMRL:axiom}.

Note that a process can only hold an endpoint of a channel (rather
than a channel per se).  Also note that each endpoint is a linear
value and it can be consumed by a function call in the sense that the
endpoint becomes unavailable for any subsequent use.

\def\tact{\mbox{\it act}}
\def\tmsg{\mbox{\bf msg}}
\def\tsend{\mbox{\bf send}}
\def\trecv{\mbox{\bf recv}}
\def\tbcast{\mbox{\bf bcast}}
\def\fchansync{\mbox{\tt chan\_sync}}
\subsection%
{Primitive formulas}
Let us use $\tact$ for both a primitive formula in LMRL and a
primitive session type.  For each $\tact$, there is a function
$\fchansync$ of the following type:
$$\fchansync:\tchan(R, \tact) \timp \tunit$$
where $\tunit$ refers to the standard unit type.  Assume that a
channel $\CH$ of some session type $\tact$ consists of $n$ endpoints
$\CH_{R_1},\ldots,\CH_{R_n}$ and each endpoint is held by one process.
When the $n$ (distinct) processes are calling $\fchansync$
simultaneously on the endpoints $\CH_{R_i}$ (for $i=1,\ldots,n$), some
actions specified by $\tact$ happen with respect to the role sets
$R_i$. It is certainly possible that the actions specified by a
particular $\tact$ can take place asynchronously, but there is no
attempt to formally address this issue here.

As an example, let $\tsend(r)$ be a primitive formula for
any given role $r$. Assume that $\CH$ is of the session type
$\tsend(r)$. Then the specified action by a call of $\fchansync$
on an endpoint $\CH_{R}$ can be described as follows:
\begin%
{itemize}
\item
Assume $r\in{R}$.
Then $\fchansync(\CH_{R})$ consumes $\CH_{R}$ after
broadcasting a message to the rest of the endpoints of $\CH$.
\item
Assume $r\not\in{R}$.
Then $\fchansync(\CH_{R})$ consumes $\CH_{R}$ after
receiving a message (sent from the endpoint $\CH_{R'}$ for the only
$R'$ containing $r$).
\end{itemize}
Clearly, we may also have a primitive formula $\trecv(r)$ for
any given role $r$. The action specified by $\trecv(r)$ is
opposite to what is specified by $\tsend(r)$: The endpoint
$\CH_{R}$ for the only $R$ containing $r$ receives a message
from each of the other endpoints.
  
As another example, let $\tmsg(r_0, r_1)$ be a primitive formula for any
given pair of distinct roles $r_0$ and $r_1$.  Assume that $\CH$ is of
the session type $\tmsg(r_0,r_1)$.  Then the specified action by a
call of $\fchansync$ on $\CH_{R}$ can be described as follows:
\begin%
{itemize}
\item
Assume $r_0\in{R}$ and $r_1\in{R}$.
Then $\fchansync(\CH_{R})$ simply consumes $\CH_{R}$.
\item
Assume $r_0\in{R}$ and $r_1\not\in{R}$.
Then $\fchansync(\CH_{R})$ consumes $\CH_{R}$ after sending a message to the
endpoint $\CH_{R'}$ for the only $R'$ containing $r_1$.
\item
Assume $r_0\not\in{R}$ and $r_1\in{R}$.  Then $\fchansync(\CH_{R})$
consumes $\CH_{R}$ after receiving a message (sent from the endpoint
$\CH_{R'}$ for the only $R'$ containing $r_0$).
\item
Assume $r_0\not\in{R}$ and $r_1\not\in{R}$.
Then $\fchansync(\CH_{R})$ simply consumes $\CH_{R}$.
\end{itemize}
In other words, $\tmsg(r_0, r_1)$ specifies a form of point-to-point
messaging from the endpoint $\CH_{R}$ to the endpoint $\CH_{R'}$
for $R$ and $R'$ containing $r_0$ and $r_1$, respectively.

\def\tneg{\mbox{\bf neg}}
\def\fchanneg{\mbox{\tt chan\_neg}}
\subsection%
{Generalized Negation}
Given an endomorphism $f$ on $\fullset$, there is a unary connective
$\lneg_{f}$ in $\LMRL$ that generalizes the notion of negation.
The meaning of $\neg_{f}(A)$ as a session type is given by the following
function for eliminating the session type constructor $\lneg_{f}$:
$$
\begin%
{array}{lcl}
\fchanneg & : & \tchan(R, \lneg_{f}(A)) \timp \tchan(\preimg{f}{R}, A) \\
\end{array}  
$$
Given an endpoint $\CH_{R}$ of type $\tchan(R, \lneg_{f}(A))$,
$\fchanneg$ turns this endpoint into one of type
$\tchan(\preimg{f}{R}, A)$.  In other words, $\lneg_{f}$ means that
the process holding an endpoint $\CH_{R}$ needs to change the role set
$R$ attached to the endpoint into the role set $\preimg{f}{R}$. In the
case where $f$ is a permutation, $\lneg_{f}$ simply means for a
process to permute according to $f$ the roles it plays (e.g., server
and client switch roles).

\subsection%
{Additive conjunction/disjunction}
\def\fchanadisj{\mbox{\tt chan\_adisj}}
\def\fchanaconjl{\mbox{\tt chan\_aconj\_l}}
\def\fchanaconjr{\mbox{\tt chan\_aconj\_r}}
\def\fchanmconj{\mbox{\tt chan\_mconj}}
\def\fchanmdisj{\mbox{\tt chan\_mdisj}}
\def\fchanmdisjl{\mbox{\tt chan\_mdisj\_l}}
\def\fchanmdisjr{\mbox{\tt chan\_mdisj\_r}}
\def\fchanappend{\mbox{\tt chan\_append}}
Given a role $r$, there is a binary connective $\aconj_{\U}$ in
$\LMRL$ for the principal ultrafilter $\U$ at $r$ (that consists of
all of the role sets containing $r$). Let us write $\aconj_{r}$ for
this $\aconj_{\U}$. Also, we may write $\tchan(R, A\aconj B)$ to mean
$\tchan(R, A\aconj_{r} B)$ for some $r\in R$ and $\tchan(R, A\adisj
B)$ to mean $\tchan(R, A\aconj_{r} B)$ for some $r\not\in R$. For each
$\CH$ of session type $A\aconj_{r} B$, there is exactly one endpoint
of type $\tchan(A\aconj B)$ and each of the other endpoints is of type
$\tchan(A\adisj B)$.
\emph{
Intuitively, a process holding an endpoint of
type $\tchan(R,A\aconj B)$ can issue an order to turn the type of the
endpoint into either $\tchan(R,A)$ or $\tchan(R,B)$ while any process
holding an endpoint of type $\tchan(R,A\adisj B)$ turns the type of
the endpoint into either $\tchan(R,A)$ or $\tchan(R,B)$ by following
an issued order.
}

Given two (linear) types $T_1$ and $T_2$, let us write $T_1\oplus T_2$
for the linear sum type of $T_1$ and $T_2$.  The following functions
are for eliminating the session type constructor $\aconj_{r}$:
$$
\begin%
{array}{lcl}
\fchanaconjl &:& \tchan(R, A\aconj B) \timp \tchan(R, A) \\
\fchanaconjr &:& \tchan(R, A\aconj B) \timp \tchan(R, B) \\
\fchanadisj  &:& \tchan(R, A\adisj B) \timp \tchan(R, A)\oplus\tchan(R, B) \\
\end{array}
$$
Assume that a channel $\CH$ is of session type $A\aconj_{r} B$ and it
consists of $n$ endpoints of the following types:
$\tchan(R_1,A\aconj_{r} B),\ldots,\tchan(R_n, A\aconj_{r} B)$.
Without loss of generality, we may assume that $r\in R_1$.  If there
are one process calling $\fchanaconjl$ ($\fchanaconjr$) on $\CH_{R_1}$
and $n-1$ processes calling $\fchanadisj$ on $\CH_{R_i}$ for
$i=2,\ldots,n$, then these calls all return; the one calling
$\fchanaconjl$ ($\fchanaconjr$) obtains the same $\CH_{R_1}$ but its
type changes to $\tchan(R_1,A)$ ($\tchan(R_1,B)$); each of the other
$n-1$ processes obtains a value of the type
$\tchan(R_i,A)\oplus\tchan(R_i,B)$ that is formed by attaching a tag
to $\CH_{R_i}$ to indicate whether $\CH_{R_i}$ is given the type
$\tchan(R_i, A)$ or $\tchan(R_i, B)$.  A direct implementation can
simply be requiring the process calling $\fchanaconjl$
($\fchanaconjr$) on $\CH_{R_1}$ to send (via the underlying physical
channel for $\CH$) the tag 0 (1) to the other processes calling
$\fchanadisj$ on $\CH_{R_i}$ for $i=2,\ldots,n$.

It should be noted that the meaning of $\aconj/\adisj$ based on a
reading of $\tchan(A\aconj B)/\tchan(A\adisj B)$ is eliminatory
(rather than introductory).  Basically, the introductory meaning of a
type $T$ is based on how a value of the type can be formed while the
eliminatory meaning of the type is based on how a value of the type
can be used.  For instance, assume that $T$ is a sum type
$T_1\oplus T_2$; a value of $T$ can be formed by either left-injecting
a value of type $T_1$ or right-injecting a value of type $T_2$; a
value of $T$ can be eliminated by performing case analysis on the
value.  Introductorily, $A\aconj B$ means to offer choice $A$ and
choice $B$ while $A\adisj B$ means to choose either $A$ or $B$.

\subsection%
{Multiplicative conjunction/disjunction}
\label{subsection:mconj/mdisj}    
Given a role $r$, there is a binary connective $\mconj_{\U}$ in
$\LMRL$ for the principal ultrafilter $\U$ at $r$. Let us write
$\mconj_{r}$ for this $\mconj_{\U}$.  Also, we may write $\tchan(R,
A\mconj B)$ to mean $\tchan(R, A\mconj_{r} B)$ for some $r\in R$ and
$\tchan(R, A\mdisj B)$ to mean $\tchan(R, A\mconj_{r} B)$ for some
$r\not\in R$. For each $\CH$ of session type $A\mconj_{r} B$, there is
exactly one endpoint of type $\tchan(A\mconj B)$ and each of the other
endpoints is of type $\tchan(A\mdisj B)$.
\emph{%
Intuitively, a process holding an endpoint of type $\tchan(R, A\mconj
B)$ turns it into two endpoints of types $\tchan(R, A)$ and $\tchan(R,
B)$ for being used in any interleaving order while any process
holding an endpoint of type $\tchan(R, A\mdisj B)$ turns it into two
endpoints of types $\tchan(R, A)$ and $\tchan(R, B)$ for being used
concurrently.  In other words, a process holding an endpoint of type
$\tchan(R, A\mconj B)$ can choose to interleave the interactions
specified by $A$ and $B$ in any order while any process holding an
endpoint of type $\tchan(R, A\mdisj B)$ must be able to handle any
chosen order of interleaving of interactions specified by $A$ and $B$.
}

Given two (linear) types $T_1$ and $T_2$, let us write $T_1\otimes
T_2$ for the linear product type of $T_1$ and $T_2$.  We can introduce
the following functions for eliminating the session type constructor
$\mconj_{r}$:
$$
\begin%
{array}{lcl}
\fchanmconj &:& \tchan(R, A\mconj B) \timp \tchan(R, A)\otimes\tchan(R, B) \\
\fchanmdisjl &:& (\tchan(R, A\mdisj B), \tchan(R, B) \timp \tunit) \timp \tchan(R, A) \\
\fchanmdisjr &:& (\tchan(R, A\mdisj B), \tchan(R, A) \timp \tunit) \timp \tchan(R, B) \\
\end{array}
$$
Assume that a channel $\CH$ is of session type $A\mconj_{r} B$ and it
consists of $n$ endpoints of the following types:
$\tchan(R_1,A\mconj_{r} B),\ldots,\tchan(R_n, A\mconj_{r} B)$.
Without loss of generality, we may assume that $r\in R_1$.  If there
are one process calling $\fchanmconj$ on $\CH_{R_1}$ and $n-1$
processes calling either $\fchanmdisjl$ or $\fchanmdisjr$ on
$\CH_{R_i}$ for $i=2,\ldots,n$ and some functions, then these calls
all return; the one calling $\fchanmconj$ on $\CH_{R_1}$ returns a
pair of endpoints $\CH'_{R_1}$ and $\CH''_{R_1}$ of types
$\tchan_{R_1}(A)$ and $\tchan_{R_1}(B)$, respectively; each of those
processes calling $\fchanmdisjl$ on $\CH_{R_i}$ (for some $i$) and
some function obtains a pair of endpoints $\CH'_{R_i}$ and
$\CH''_{R_i}$ and then keeps $\CH'_{R_i}$ as the return value while
passing $\CH''_{R_i}$ to the function and \emph{initiating a thread to
  execute the function call}; each of those processes calling
$\fchanmdisjr$ does analogously.

We do not specify the manner in which $\CH'$ and $\CH''$ are actually
created or obtained as there are many possibilities in practice. For
instance, it is possible to use $\CH$ for $\CH''$ after requiring the
process holding the endpoint $\CH_{R_1}$ to create a fresh channel
$\CH'$ so that the process keeps the endpoint $\CH'_{R_1}$ for itself
and sends each $\CH'_{R_i}$ (via $\CH$) to the process holding
$\CH_{R_i}$, where $i$ ranges from $2$ to $n$. With this
implementation strategy, $A\mconj B$ ($A\mdisj B$) is often
interpreted as \emph{output $A$ and then behave as $B$} (\emph{input
  $A$ and then behave as $B$})~\cite{Wadler12-Prop-as-Sessions}.
However, we see this interpretation as a form of convenience (rather
than a logical consequence derived from $\LMRL$). For instance, we can
certainly have a different implementation that chooses a particular
$r$ (e.g., $0$) and requires the process holding $\CH_{R}$ for the
only $R$ containing $r$ to create $\CH'$ and then sends its endpoints
elsewhere. As far as we can see, $\LMRL$ specifies in the case of
$\mconj_r$ whether the two obtained endpoints $\CH'_{R}$ and
$\CH''_{R}$ should be used interleavingly or concurrently but it does
not specify how these two endpoints are actually obtained.

\subsection%
{Session concatenation}
\def\mappend{@}
Given two session types $A$ and $B$, a process holding an endpoint of
type $\tchan(R, A\mconj_r B)$ for some $r\in R$ can turn it into two
endpoints of types $\tchan(R, A)$ and $\tchan(R, B$) and then use them
in any interleaving order it desires. We can have $\mappend_r$ as a
variant of $\mconj_r$: A process holding an endpoint of type
$\tchan(R, A\mappend_r B)$ for some $r\in R$ can turn it into two
endpoints of types $\tchan(R, A)$ and $\tchan(R, B$) but is required
to finish using the one of type $\tchan(R, A)$ before starting to use
the other. There is actually no need in this case for the process to
hold two endpoints simultaneously: We can introduce the following
function $\fchanappend$ that uses an endpoint of type $\tchan(R,
A\mappend_r B)$ first as an endpoint of type $\tchan(R, A)$ and then
as an endpoint of type $\tchan(R, B)$:
$$
\begin%
{array}{lcl}
\fchanappend &:& (\tchan(R, A\mappend B), \tchan(R, A) \timp T) \timp T\otimes\tchan(R, B) \\
\end{array}
$$
Note that $A\mappend B$ (instead of $A\mappend_r B$) occurs in the
type of $\fchanappend$. Unlike $\fchanmconj$ (for $\mconj_r$),
$\fchanappend$ is not required to send new endpoints to other
processes and thus there is no difference between $\mappend_{r}$ and
$\mappend_{r'}$ even if $r$ and $r'$ are distinct. In essence, the
function $\fchanappend$ takes an endpoint $\CH_{R}$ of type $\tchan(R,
A\mappend B)$ and a function; it first treats the endpoint $\CH_{R}$
as one of type $\tchan(R, A)$ and calls the function on it; the value
returned by the function call is then paired with $\CH_R$ as the
return value (of the call to $\fchanappend$).

\def\fchansend{\mbox{\tt chan\_send}}
\def\fchanrecv{\mbox{\tt chan\_recv}}
Given two distinct roles $r_0$ and $r_1$, we used $\tmsg(r_0, r_1)$ to
mean messaging from $r_0$ to $r_1$. We now use $\tmsg(r_0, r_1, T)$ to
mean a value of type $T$ sent from $r_0$ to $r_1$. We can introduce
the following function for sending:
$$
\begin%
{array}{lcl}
\fchansend &:& (\tchan(R, \tmsg(r_0, r_1, T)\mappend A), T)\timp \tchan(R, A) \\
\end{array}
$$
where it is assumed that $r_0\in R$ and $r_1\not\in R$, and the following
function for receiving:
$$
\begin%
{array}{lcl}
\fchanrecv &:& (\tchan(R, \tmsg(r_0, r_1, T)\mappend A))\timp T\otimes\tchan(R, A) \\
\end{array}
$$
where it is assumed that $r_0\not\in R$ and $r_1\in R$.  In the case
where there are only two roles $0$ and $1$, we may choose to write
$\tsend(T)$ for $\tmsg(0,1,T)$ and $\trecv(T)$ for $\tmsg(1,0,T)$.  If
one desires, one can even write $T\mconj A$ for $\tsend(T)\mappend A$
and $T\mdisj A$ for $\trecv(T)\mappend A$ (or $T\mconj A$ for
$\trecv(T)\mappend A$ and $T\mdisj A$ for $\tsend(T)\mappend A$).
In this case,
it is fine to interpret $\otimes/\invamp$ as output/input, and it is
also fine to interpret $\otimes/\invamp$ as input/output.

\subsection%
{More session constructors}
\label{subsection:more_session_constructors}
We may introduce a primitive formula $\tnone$ for which the specified
action is simply doing nothing. Given a role $r$ and a session type
$A$, we define $\toption(r,A)$ as $A\aconj_{r}\tnone$ and define
$\trepseq(r, A)$ recursively as $\toption(r, A@\trepseq(r, A))$.
Assume $r\in R$. A process holding an endpoint $\CH_R$ of type
$\tchan(R, \toption(r, A))$ can choose to perform a session specified
by $A$ on $\CH_R$ or simply terminate the use of $\CH_R$.  A process
holding an endpoint $\CH_R$ of type $\tchan(R, \trepseq(r, A))$ can
choose to perform a session specified by $A$ repeatedly.

{\bf Example 1.}~~~%
Let us assume the availability of 3 roles: Seller(0), Buyer1(1) and
Buyer2(2). A description of the one-seller-and-two-buyers (S0B1B2)
protocol due to~\cite{HondaYC08} can essentially be given as follows:
{Buyer1 sends a book title to Seller; Seller replies a price quote to
both Buyer1 and Buyer2; Buyer1 tells Buyer2 how much he can
contribute; if Buyer2 is willing to pay for the remaining part, she
sends Seller a proof of payment and Seller sends her a receipt for the
sale (and the session ends); if Buyer2 is not willing, she terminates
the session.}
Following is an encoding of the S0B1B2 protocol as a session
type:
\begin%
{center}
\begin%
{minipage}{10cm}
\begin%
{verbatim}
title(1, 0)@quote(0, 1)@quote(0, 2)@
contrib(1, 2)@option(2, proof(2, 0)@receipt(0, 2))
\end{verbatim}
\end{minipage}\end{center}\vspace{2pt}
where \verb`title(1, 0)` means sending the title from role 1 to role 0;
\verb`quote(0, 1)` means sending the price quote from role 0 to role
1; \verb`contrib(1, 2)` means sending the amount of contribution of
Buyer1 from role 1 to role 2; etc.

{\bf Example 2.}~~~%
Let us assume the availability of three roles: Client(0), Server(1),
and Verifier(2).  A protocol for login involving three parties can be
described as follows: Client sends userid to Server; Server sends the
received userid to Verifier; Verifier queries Client an indefinite
number of times; Verifier sends the verification result to Server.
Following is an encoding of this protocol as a session type:
\begin%
{center}
\begin%
{minipage}{9cm}
\begin%
{verbatim}
userid(0,1)@userid(1,2)@
repseq(2,query(2,0)@answer(0,2))@result(2,1)
\end{verbatim}
\end{minipage}\end{center}\vspace{2pt}
where
\verb`userid(0,1)` and \verb`userid(1,2)`
mean sending a userid from role 0 to role 1
and from role 1 to role 2, respectively;
\verb`query(2,0)` means sending a query question from role 2 to role 0;
\verb`answer(0,2)` means sending an answer (to the received question)
from role 0 to role 2; \verb`result(2,1)` means sending a result from
role 2 to role 1 (to indicate success or failure of verification).

\def\tscore{\mbox{\it score}}
\def\tanswer{\mbox{\it answer}}
{\bf Example 3.}~~~%
Let us assume the availability of three roles: Judge(0), Contestant1(1),
and Contestant2(2). A protocol for some kind of contest involving three
parties can be described as follows: Judge broadcasts a query to
Contestant1 and Contestant2; each contestant sends his or her answer
to Judge and then receives a score from Judge.
Following is an encoding of this protocol as a session type:
\begin%
{center}
\begin%
{minipage}{12cm}
\begin%
{verbatim}
query(0)@(mconj(0, answer(1,0)@score(0,1), answer(2,0)@score(0,2)))
\end{verbatim}
\end{minipage}\end{center}\vspace{2pt}
where \verb`query(0)` means broadcasting a query from role 0 to all of
the other roles; the syntax \verb`mconj(0, ...)` means $\mconj_{0}$;
\verb`answer(1,0)` means sending an answer from role 1 to role 0 and
\verb`score(0,1)` means sending a score from role 0 to role 1;
\verb`answer(2,0)` and \verb`score(0,2)` mean similarly.  Let $A_i$ be
$\tanswer(i,0)@\tscore(0,i)$ for $i=1,2$.  At some moment, Judge holds two
endpoints of types $\tchan(\myset{0},A_1)$ and
$\tchan(\myset{0},A_2)$, and he can certainly use them in two
concurrently running threads (but is not required to do so).  On the
other hand, Contestant1 holds two endpoints of types
$\tchan(\myset{1},A_1)$ and $\tchan(\myset{1},A_2)$ at the same moment
and is required to use them concurrently. And the same can be said about
Contestant2.

\subsection%
{Splitting an endpoint}
\def\fchansplit{\mbox{\tt chan\_split}}
Lemma~\ref{lemma:LMRL:roleset-split} corresponds to a function for
splitting a given endpoint:
$$
\begin%
{array}{lcr}
\fchansplit & : & (\tchan(R_1\uplus{R}_2,A), \tchan(R_1, A)\timp 1)\timp\tchan(R_2, A)
\end{array}
$$
When applied to an endpoint $\CH_{R_1\uplus{R}_2}$ and a function, $\fchansplit$
\emph{initiates a thread} to call the function
on the endpoint (for it being used as $\CH_{R_1}$)
and then returns the endpoint (for it to be used as $\CH_{R_2}$).

In the context of session-typed multiparty channels, the
inadmissibility of the following inference rule (for disjoint $R_1$
and $R_2$) should be clear:
$$
(\interp{R_1}{A}, \interp{R_2}{A}) \tpjg \interp{R_1\uplus{R}_2}{A}
$$
We cannot simply merge two given endpoints $\CH'_{R_1}$ and
$\CH''_{R_2}$ as $\CH'$ and $\CH''$ may not be the same channel. If
they happen to be the same channel, then allowing a process to hold
both $\CH'_{R_1}$ and $\CH''_{R_2}$ can potentially lead to
deadlocking.

\subsection%
{Multiparty cut-elimination}
\label{subsection:multiparty-cut-elimination}    
\def\fchanonecut{\mbox{\tt chan\_1\_cut}}
\def\fchantwocut{\mbox{\tt chan\_2\_cut}}
\def\fchanthreecut{\mbox{\tt chan\_3\_cut}}
\def\fchantwocutres{\mbox{\tt chan\_2\_cutres}}
Lemma~\ref{lemma:LMRL:1-cut} corresponds to the following function
that can be called to eliminate any endpoint $\CH_{\emptyset}$:
$$
\begin%
{array}{lcr}
\fchanonecut & : & (\tchan(\emptyset,A)) \timp \tunit
\end{array}
$$
Lemma~\ref{lemma:LMRL:mp-cut} (or more precisely, its proof) indicates
the existence of a generic method for forwarding messages between three
endpoints of certain matching types:
$$
\begin%
{array}{lcr}
\fchanthreecut & : & (\tchan(R_1,A), \tchan(R_2,A), \tchan(R_3,A)) \timp \tunit
\end{array}
$$
where $\setcomp{R}_1\uplus\setcomp{R}_2\uplus\setcomp{R}_3=\fullset$
is assumed and each endpoint belongs to a distinct channel.  The
following $\fchantwocut$ can be seen a special case of
$\fchanthreecut$ where $R_1=R$, $R_2=\setcomp{R}$ and $R_3=\fullset$:
$$
\begin%
{array}{lcr}
\fchantwocut & : & (\tchan(R,A), \tchan(\setcomp{R},A)) \timp \tunit
\end{array}
$$

When applied to three endpoints $\CH^{i}_{R_i}$ for $1\leq i\leq 3$,
$\fchanthreecut$ behaves in the following manner: Suppose that a message
for role $r$ is received at $\CH^{1}_{R_1}$; we have
$r\in\setcomp{R}_2\uplus\setcomp{R}_3$ as $r\in R_1$ holds; if
$r\in\setcomp{R}_2$, the received message is sent via the endpoint
$\CH^2_{R_2}$ to the process holding $\CH^2_{R}$ for some $R$
satisfying $r\in R$; if $r\in\setcomp{R}_3$, the received message is
sent via the endpoint $\CH^3_{R_3}$ to the process holding $\CH^3_{R}$
for some $R$ satisfying $r\in R$.
\emph%
{After $\fchanthreecut$ is called on
$\CH^1_{R_1}$, $\CH^2_{R_2}$ and $\CH^3_{R_3}$, all of the other
endpoints of $\CH^1$, $\CH^2$ and $\CH^3$ are considered belonging to
one single channel}.
Implementing $\fchanthreecut$ consists of a crucial step in implementation
of multiparty channels.

\subsection%
{Modality and Quantification}    
The exponentials $\bang/\qmark$ in linear logic
are modal connectives.
Given a role $r$, there is a connective $\bang_{\U}$ in $\LMRL$ for
the principal ultrafilter $\U$ at $r$. Let us write $\bang_{r}$ for
this $\bang_{\U}$. Essentially, $\bang_{r}(A)$ for any given $A$ can
be replaced with $\trepeat(r, A)$, which is recursively defined as
$A\mconj_{r}\trepeat(r, A)$. As a session type, $\bang_{r}(A)$ means
that a sequence of interactions specified by $A$ can be repeated
concurrently (while $\trepseq(r, A)$ means sequential repetition).

Given a role $r$, there is also a quantifier $\forall_{\U}$ for in
$\LMRL$ for the principal ultrafilter $\U$ at $r$.  Let us write
$\forall_{r}$ for this $\forall_{\U}$, which is supposed to be used
for constructing dependent session types (that are not studied here).

\def\rc{\mbox{\it rc}}
\def\ccon{\mbox{\it cc}}
\def\cfun{\mbox{\it cf}}
\def\ctrue{\mbox{\it true}}
\def\cfalse{\mbox{\it false}}
\def\MTLCZ{\mbox{MTLC}_0}
\def\MTLCO{\mbox{MTLC}_1}
\def\ty{\mbox{$T$}}
\def\vwty{\mbox{$\hat{T}$}}
\def\bty{\mbox{$\delta$}}
\def\bvwty{\mbox{$\hat{\delta}$}}
\def\tint{\mbox{\bf int}}
\def\tbool{\mbox{\bf bool}}
\def\itimp{\rightarrow_{i}}
\def\ltimp{\rightarrow_{l}}
\def\cst{c}
\def\exp{e}
\def\vexp{\vec{\exp}}
\def\val{v}
\def\vval{\vec{\val}}
\def\dif{\mbox{\tt if}}
\def\dfst#1{\mbox{\tt fst}(#1)}
\def\dsnd#1{\mbox{\tt snd}(#1)}
\def\dunit{\langle\rangle}
\def\tuple#1{\langle#1\rangle}
\def\lam#1#2{{\tt lam}\;#1.\,#2}
\def\app#1#2{{\tt app}(#1, #2)}
\def\fix#1#2{{\tt fix}\;#1.\,#2}
\def\letin#1#2{{\tt let}\;#1\;{\tt in}\;#2\;{\tt end}}
\section%
{$\MTLCZ$:
 Linearly typed Multi-threaded $\lambda$-calculus}
\label{section:MTLCZ}
\begin%
{figure}
\[%
\begin%
{array}{lrcl}
\mbox{expr.} & \exp & ::= &
x \mid f \mid \rc \mid \cst(\vexp) \mid \\
             &      &     &
\dunit \mid \tuple{\exp_1,\exp_2} \mid \dfst{\exp} \mid \dsnd{\exp} \mid \\
             &      &     &
\letin{\tuple{x_1,x_2}=\exp_1}{\exp_2} \mid \\
             &      &     &
\lam{x}{\exp} \mid \app{\exp_1}{\exp_2} \mid \fix{x}{\val}  \\
\mbox{values} & v & ::= &
x \mid \rc \mid \ccon(\vval) \mid \dunit \mid \tuple{\val_1,\val_2} \mid \lam{x}{\exp} \\
\mbox{types} & \ty & ::= &
\bty \mid \tunit \mid \ty_1*\ty_2 \mid \vwty_1\itimp\vwty_2  \\
\mbox{viewtypes} & \vwty & ::= &
\bvwty \mid \ty \mid \vwty_1\otimes\vwty_2 \mid \vwty_1\ltimp\vwty_2 \\
\end{array}\]
\caption{Some syntax for $\MTLCZ$}
\label{figure:MTLCZ:syntax}
\end{figure}
While what is presented in Section~\ref{section:LMRL_session_types}
may sound intuitive, it is not formal. In this section and
the next, we formulate a design of linearly typed channels for
multiparty communication where the types are directly rooted in
$\LMRL$.  This formulation itself is standard and the contribution it
brings to this paper is mostly technical.

We first present a multi-threaded lambda-calculus $\MTLCZ$ equipped
with a simple linear type system, setting up the basic machinery for
further development. Some syntax of $\MTLCZ$ is given in
Figure~\ref{figure:MTLCZ:syntax}. We use $\vexp$ and $\vval$ for
sequences of expressions and values, respectively. We use $\rc$ for
constant resources and $\cst$ for constants, which include both
constant constructors $\ccon$ and constant functions $\cfun$. We treat
resources in $\MTLCZ$ abstractly and will later introduce
communication channels as a concrete form of resources. The meaning of
various standard forms of expressions in $\MTLCZ$ should be
intuitively clear. We may refer to a closed expression (containing no
free variables) as a program.

We use $\ty$ and $\vwty$ for (non-linear) types and (linear)
viewtypes, respectively.  Note that a type is always considered a
viewtype and a viewtype is referred to as a true viewtype if it is not
a type. We use $\bty$ and $\bvwty$ for base types and base viewtypes,
respectively. For instance, $\tint$ and $\tbool$ are base types for
integers and booleans, respectively. We also assume the availability
of integer constants and sets of integer constants for forming types.
For instance, we may have a singleton type $\tint(i)$ for each integer
$i$, which can only be assigned to a dynamic expression of value equal
to $i$.

\def\SIG{\mbox{\it SIG}}
\def\ctimp{\Rightarrow}
\def\iadd{\mbox{\it iadd}}
For a simplified presentation, we do not introduce any concrete base
viewtypes in $\MTLCZ$. We assume a signature $\SIG$ for assigning a
viewtype to each constant resource $\rc$ and a constant type (c-type)
schema of the form $(\vwty_1,\ldots,\vwty_n)\ctimp\vwty_0$ to each
constant. For instance, we may have a constant function $\iadd$
of the following c-type schema:
$$
\begin%
{array}{rcl}
\iadd & : & (\tint(i), \tint(j)) \ctimp \tint(i+j)
\end{array}
$$
where $i$ and $j$ are meta-variables ranging over integer constants.
Each occurrence of $\iadd$ in a program is given a c-type that is an
instance of the c-type schema assigned to $\iadd$.

Let $\vwty_1$ and
$\vwty_2$ be two viewtypes. The type constructor $\otimes$ is based on
multiplicative conjunction in linear logic.  Intuitively, if a
resource is assigned the viewtype $\vwty_1\otimes\vwty_2$, then the
resource is a conjunction of two resources of viewtypes $\vwty_1$ and
$\vwty_2$.  The type constructor $\ltimp$ is essentially based on
linear implication $\multimap$ in linear logic.  Given a function of
the viewtype $\vwty_1\ltimp\vwty_2$ and a value of the viewtype
$\vwty_1$, applying the function to the value yields a result of the
viewtype $\vwty_2$ while the function itself is consumed. If the
function is of the type $\vwty_1\itimp\vwty_2$, then applying the
function does not consume it. The subscript $i$ in $\itimp$ is often
dropped, that is, $\timp$ is assumed to be $\itimp$ by default.  The
meaning of various forms of types and viewtypes is to be made clear
and precise when the rules are presented for assigning viewtypes to
expressions in $\MTLCZ$.

\def\dom{\mbox{\bf dom}}
\def\mapadd#1#2#3{#1[#2\mapsto #3]}
\def\mapdel#1#2{#1\backslash#2}
\def\maprep#1#2#3{#1[#2 := #3]}
\def\fthreadcreate{\mbox{\it thread\_create}}
There is a special constant function $\fthreadcreate$ in $\MTLCZ$ for
thread creation, which is assigned the following interesting c-type:
\[\begin{array}{rcl}
\fthreadcreate & : & (\tunit\ltimp\tunit) \ctimp \tunit
\end{array}\]
A function of the type $\tunit\ltimp\tunit$ is a procedure that takes no
arguments and returns no result (when its evaluation terminates).  Given
that $\tunit\ltimp\tunit$ is a true viewtype, a procedure of this type may
contain resources and thus must be called exactly once. The operational
semantics of $\fthreadcreate$ is to be formally defined later.

A variety of mappings, finite or infinite, are to be introduced in the
rest of the presentation.  We use $[]$ for the empty mapping and
$[i_1,\ldots,i_n\mapsto o_1,\ldots,o_n]$ for the finite mapping that maps
$i_k$ to $o_k$ for $1\leq k\leq n$.  Given a mapping $m$, we write
$\dom(m)$ for the domain of $m$. If $i\not\in\dom(m)$, we use
$\mapadd{m}{i}{o}$ for the mapping that extends $m$ with a link from $i$ to
$o$.  If $i\in\dom(m)$, we use $\mapdel{m}{i}$ for the mapping obtained
from removing the link from $i$ to $m(i)$ in $m$, and $\maprep{m}{i}{o}$
for $\mapadd{(\mapdel{m}{i})}{i}{o}$, that is, the mapping obtained from
replacing the link from $i$ to $m(i)$ in $m$ with another link from $i$ to
$o$.

\def\resof{\rho}
\def\RES{{\bf RES}}
\def\rns{{\mathcal R}}
\begin{figure}
\[\begin{array}{rcl}
\resof(\rc) & = & \{\rc\} \\
\resof(\cst(\exp_1,\ldots,\exp_n)) & = & \resof(\exp_1)\uplus\cdots\uplus\resof(\exp_n) \\
\resof(x) & = & \emptyset \\
\resof(\dunit) & = & \emptyset \\
\resof(\tuple{\exp_1,\exp_2}) & = & \resof(\exp_1)\uplus\resof(\exp_2) \\
\resof(\dfst{\exp}) & = & \resof(\exp) \\
\resof(\dsnd{\exp}) & = & \resof(\exp) \\
\resof(\dif(\exp_0,\exp_1,\exp_2)) & = & \resof(\exp_0)\uplus\resof(\exp_1) \\
\resof(\letin{\tuple{x_1,x_2}=\exp_1}{\exp_2}) & = & \resof(\exp_1)\uplus\resof(\exp_2) \\
\resof(\lam{x}{\exp}) & = & \resof(\exp) \\
\resof(\app{\exp_1}{\exp_2}) & = & \resof(\exp_1)\uplus\resof(\exp_2) \\
\resof(\fix{x}{\val}) & = & \resof(\val) \\
\end{array}\]
\caption{The definition of $\resof(\cdot)$}
\label{figure:resof}
\end{figure}
We define a function $\resof(\cdot)$ in Figure~\ref{figure:resof} to
compute the multiset (that is, bag) of constant resources in a given
expression.  Note that $\uplus$ denotes the multiset union. In the type
system of $\MTLCZ$, it is to be guaranteed that $\resof(\exp_1)$ equals
$\resof(\exp_2)$ whenever an expression of the form
$\dif(\exp_0,\exp_1,\exp_2)$ is constructed, and this justifies
$\resof(\dif(\exp_0,\exp_1,\exp_2))$ being defined as
$\resof(\exp_0)\uplus\resof(\exp_1)$.

We use $\rns$ to range over finite multisets of resources. Therefore,
$\rns$ can also be regarded as a mapping from resources to natural numbers:
$\rns(\rc) = n$ means that there are $n$ occurrences of $\rc$ in
$\rns$. It is clear that we may not combine resources arbitrarily. For
instance, we may want to exclude the combination of one resource stating
integer 0 at a location L and another one stating integer 1 at the same
location. We fix an abstract collection $\RES$ of finite multisets of
resources and assume the following:
\begin{itemize}
\item
$\emptyset\in\RES$.
\item
For any $\rns_1$ and
$\rns_2$, $\rns_2\in\RES$ if $\rns_1\in\RES$ and $\rns_2\subseteq\rns_1$,
where $\subseteq$ is the subset relation on multisets.
\end{itemize}
We say that $\rns$ is a valid multiset of resources if $\rns\in\RES$ holds.

\def\pool{\Pi}
\def\tid{\mbox{\it tid}}
In order to formalize threads, we introduce a notion of {\em
  pools}. Conceptually, a pool is just a collection of programs (that is,
closed expressions).  We use $\pool$ for pools, which are formally defined
as finite mappings from thread ids (represented as natural numbers) to
(closed) expressions in $\MTLCZ$ such that $0$ is always in the domain of
such mappings.  Given a pool $\pool$ and $\tid\in\dom(\pool)$, we refer to
$\pool(\tid)$ as a thread in $\pool$ whose id equals $\tid$. In particular,
we refer to $\pool(0)$ as the main thread in $\pool$.  The definition of
$\resof(\cdot)$ is extended as follows to compute the multiset of resources
in a given pool:
\begin%
{center}
$\resof(\pool)=\biguplus_{\tid\in\dom(\pool)}\resof(\pool(\tid))$
\end{center}
We are to define a relation on pools in
Section~\ref{section:MTLCZ:dynamics} to simulate multi-threaded program
execution.

\begin{figure}
\[\begin{array}{c}
\infer[\mbox{\bf (ty-res)}]
      {\Gamma;\emptyset\tpjg\rc:\bvwty}
      {\SIG\temd\rc:\bvwty} \\[4pt]
\infer[\mbox{\bf (ty-cst)}]
      {\Gamma;\Delta_1,\ldots,\Delta_n\tpjg\cst(\exp_1,\ldots,\exp_n):\vwty}
      {
       \SIG\temd\cst:(\vwty_1,\ldots,\vwty_n)\ctimp\vwty &
       \Gamma;\Delta_i\tpjg\exp_i:\vwty_i~~\mbox{for $1\leq i\leq n$}
      } \\[4pt]
\infer[\mbox{\bf (ty-var-i)}]
      {(\Gamma,x:\ty;\emptyset)\tpjg x:\ty}{}
\kern12pt
\infer[\mbox{\bf (ty-var-l)}]
      {(\Gamma;\emptyset,x:\vwty)\tpjg x:\vwty}{} \\[4pt]
\infer[\mbox{\bf(ty-if)}]
      {\Gamma;\Delta_0,\Delta\tpjg\dif(\exp_0,\exp_1,\exp_2):\vwty}
      {
       \resof(\exp_1) = \resof(\exp_2) &
       \Gamma;\Delta_0\tpjg\exp_0:\tbool &
       \Gamma;\Delta\tpjg \exp_1:\vwty & \Gamma;\Delta\tpjg \exp_2:\vwty &
      } \\[4pt]
\infer[\mbox{\bf(ty-unit)}]
      {\Gamma;\emptyset\tpjg\dunit:\tunit}{} \\[4pt]
\infer[\mbox{\bf(ty-tup-i)}]
      {\Gamma;\Delta_1,\Delta_2\tpjg\tuple{\exp_1,\exp_2}:\ty_1*\ty_2}
      {\Gamma;\Delta_1\tpjg\exp_1:\ty_1 &
       \Gamma;\Delta_2\tpjg\exp_2:\ty_2 } \\[4pt]
\infer[\mbox{\bf(ty-fst)}]
      {\Gamma;\Delta\tpjg\dfst{\exp}:\ty_1}
      {\Gamma;\Delta\tpjg\exp:\ty_1*\ty_2}
\kern12pt
\infer[\mbox{\bf(ty-snd)}]
      {\Gamma;\Delta\tpjg\dsnd{\exp}:\ty_2}
      {\Gamma;\Delta\tpjg\exp:\ty_1*\ty_2} \\[4pt]
\infer[\mbox{\bf(ty-tup-l)}]
      {\Gamma;\Delta_1,\Delta_2\tpjg\tuple{\exp_1,\exp_2}:\vwty_1\otimes\vwty_2}
      {\Gamma;\Delta_1\tpjg\exp_1:\vwty_1 &
       \Gamma;\Delta_2\tpjg\exp_2:\vwty_2 } \\[4pt]
\infer[\mbox{\bf(ty-tup-l-elim)}]
      {\Gamma;\Delta_1,\Delta_2\tpjg\letin{\tuple{x_1,x_2}=\exp_1}{\exp_2}:\vwty}
      {\Gamma;\Delta_1\tpjg\exp_1:\vwty_1\otimes\vwty_2 &
       \Gamma;\Delta_2,x_1:\vwty_1,x_2:\vwty_2\tpjg\exp_2:\vwty} \\[4pt]
\infer[\mbox{\bf(ty-lam-l)}]
      {\Gamma;\Delta\tpjg\lam{x}{\exp}:\vwty_1\ltimp\vwty_2}
      {(\Gamma;\Delta),x:\vwty_1\tpjg\exp:\vwty_2} \\[4pt]
\infer[\mbox{\bf(ty-app-l)}]
      {\Gamma;\Delta_1,\Delta_2\tpjg\app{\exp_1}{\exp_2}:\vwty_2}
      {\Gamma;\Delta_1\tpjg\exp_1:\vwty_1\ltimp\vwty_2 &
       \Gamma;\Delta_2\tpjg\exp_2:\vwty_1 } \\[4pt]
\infer[\mbox{\bf(ty-lam-i)}]
      {\Gamma;\emptyset\tpjg\lam{x}{\exp}:\vwty_1\itimp\vwty_2}
      {(\Gamma;\emptyset),x:\vwty_1\tpjg\exp:\vwty_2 &
       \resof(\exp) = \emptyset} \\[4pt]
\infer[\mbox{\bf(ty-app-i)}]
      {\Gamma;\Delta_1,\Delta_2\tpjg\app{\exp_1}{\exp_2}:\vwty_2}
      {\Gamma;\Delta_1\tpjg\exp_1:\vwty_1\itimp\vwty_2 &
       \Gamma;\Delta_2\tpjg\exp_2:\vwty_1 } \\[4pt]
\infer[\mbox{\bf(ty-fix)}]
      {\Gamma;\emptyset\tpjg\fix{x}{\val}:\ty}
      {\Gamma, x:\ty;\emptyset\tpjg\val:\ty} \\[4pt]
\infer[\mbox{\bf(ty-pool)}]
      {\tpjg\pool:\vwty}
      {(\emptyset;\emptyset)\tpjg\pool(0):\vwty &
       (\emptyset;\emptyset)\tpjg\pool(\tid):\tunit~~\mbox{for each $0<\tid\in\dom(\pool)$}
      } \\[4pt]
\end{array}\]
\caption{The typing rules for $\MTLCZ$}
\label{figure:MTLCZ:typing_rules}
\end{figure}
\subsection{Static Semantics}
We use $\Gamma$ for a typing context that assigns (non-linear) types
to variables and $\Delta$ for a typing context that assigns (linear)
viewtypes to variables.  Any typing context can be regarded as a
finite mapping as each variable can occur at most once.  Given
$\Gamma_1$ and $\Gamma_2$ satisfying
$\dom(\Gamma_1)\cap\dom(\Gamma_2)=\emptyset$, we write
$(\Gamma_1,\Gamma_2)$ for the union of $\Gamma_1$ and $\Gamma_2$.  The
same notation also applies to linear typing contexts ($\Delta$).
Given $\Gamma$ and $\Delta$, we can form a combined typing context
$(\Gamma;\Delta)$ if $\dom(\Gamma)\cap\dom(\Delta)=\emptyset$.  Given
$(\Gamma;\Delta)$, we may write $(\Gamma;\Delta),x:\vwty$ for either
$(\Gamma;\Delta,x:\vwty)$ or $(\Gamma,x:\vwty;\Delta)$ (if $\vwty$ is
actually a type).


A typing judgment in $\MTLCZ$ is of the form
$(\Gamma;\Delta)\tpjg\exp:\vwty$, meaning that $\exp$ can be assigned
the viewtype $\vwty$ under $(\Gamma;\Delta)$.  The typing rules for
$\MTLCZ$ are listed in Figure~\ref{figure:MTLCZ:typing_rules}.
In the rule $\mbox{\bf(ty-cst)}$, the following judgment requires that
the c-type be an instance of the c-type schema assigned to $\cst$ in
$\SIG$:
$$\SIG\temd\cst:(\vwty_1,\ldots,\vwty_n)\ctimp\vwty$$

By inspecting the typing rules in
Figure~\ref{figure:MTLCZ:typing_rules}, we can readily see that a
closed value cannot contain any resources if the value itself can be
assigned a type (rather than a linear type).  More formally, we have
the following proposition:
\begin%
{proposition}
\label{proposition:value_type}
If $(\emptyset;\emptyset)\tpjg\val:\ty$ is derivable,
then $\resof(\val)=\emptyset$.
\end{proposition}
This proposition plays a fundamental role in $\MTLCZ$:
The rules in Figure~\ref{figure:MTLCZ:typing_rules} are actually
so formulated in order to make the proposition hold.

The following lemma, which is often referred to as
{\em Lemma of Canonical Forms}, relates the form of a value to its type:
\begin{lemma}\label{lemma:MTLCZ:canonical}
Assume that $(\emptyset;\emptyset)\tpjg\val:\vwty$ is derivable.
\begin{itemize}
\item
If $\vwty=\bty$,
then $\val$ is of the form $\ccon(\val_1,\ldots,\val_n)$.
\item
If $\vwty=\bvwty$,
then $\val$ is of the form $\rc$ or $\ccon(\val_1,\ldots,\val_n)$.
\item
If $\vwty=\tunit$, then $\val$ is $\dunit$.
\item
If
$\vwty=\ty_1*\ty_2$
or
$\vwty=\vwty_1\otimes\vwty_2$,
then $\val$ is of the form $\tuple{\val_1,\val_2}$.
\item
If $\vwty=\vwty_1\itimp\vwty_2$ or $\vwty=\vwty_1\ltimp\vwty_2$, then
$\val$ is of the form $\lam{x}{\exp}$.
\end{itemize}
\end{lemma}
\begin{proof}
By an inspection of the rules in Figure~\ref{figure:MTLCZ:typing_rules}.
\end{proof}

\def\esubst{\theta}
We use $\esubst$ for substitution on variables $x$:
\[\begin{array}{rcl}
\esubst & ::= & [] \mid \esubst[x\mapsto\val] \\
\end{array}\]
For each $\esubst$, we define the multiset $\resof(\esubst)$ of
resources in $\esubst$ as the union of $\resof(\esubst(x))$ for
$x\in\dom(\esubst)$.  Given an expression $\exp$, we use
$\exp[\esubst]$ for the result of applying $\esubst$ to $\exp$.  We write
$(\Gamma_1;\Delta_1)\tpjg\esubst:(\Gamma_2;\Delta_2)$ to mean the following:
\begin{itemize}
\item
$\dom(\esubst)=\dom(\Gamma_2)\cup\dom(\Delta_2)$, and
\item
$(\Gamma_1;\emptyset)\tpjg\esubst(x):\Gamma_2(x)$
is derivable for each $x\in\Gamma_2$, and
\item
there exists a linear typing context $\Delta_{1,x}$ for each
$x\in\dom(\Delta_2)$ such that
$(\Gamma_1;\Delta_{1,x})\tpjg\esubst(x):\Delta_2(x)$
is derivable, and
\item
$\Delta_1$ is the union of $\Delta_{1,x}$ for ${x\in\dom(\Delta_2)}$.
\end{itemize}
The following lemma, which is often referred to as
{\em Substitution Lemma}, is needed to establish the soundness of the type
system of $\MTLCZ$:
\begin%
{lemma}\label{lemma:MTLCZ:substitution}
Assume $(\Gamma_1;\Delta_1)\tpjg\esubst:(\Gamma_2;\Delta_2)$ and
$(\Gamma_2;\Delta_2)\tpjg\exp:\vwty$.
Then $(\Gamma_1;\Delta_1)\tpjg\exp[\esubst]:\vwty$ is derivable
and $\resof(\exp[\esubst])=\resof(\exp)\uplus\resof(\esubst)$.
\end{lemma}
\begin{proof}
By induction on the derivation of $(\Gamma_2;\Delta_2)\tpjg\exp:\vwty$.
\end{proof}

\subsection
{Dynamic Semantics}
\label{section:MTLCZ:dynamics}
The evaluation contexts in $\MTLCZ$ are defined below:
\[
\begin%
{array}{l}
\mbox{eval.~ctx.}
E~~::=~~ \\
{\kern6pt}
{[] \mid \cst(\vval,E,\vexp) \mid
 \dif(E,\exp_1,\exp_2) \mid \tuple{E,\exp} \mid \tuple{\val, E} \mid} \\
{\kern6pt}
{\letin{\tuple{x_1,x_2}=E}{\exp} \mid \dfst{E} \mid \dsnd{E} \mid \app{E}{\exp} \mid \app{\val}{E}} \\
\end{array}\]
Given an evaluation context $E$ and an expression $\exp$, we use $E[\exp]$
for the expression obtained from replacing the only hole $[]$ in $E$ with
$\exp$.

\def\esubst#1#2#3{#3[#2\mapsto #1]}
\begin%
{definition}
We define pure redexes and their reducts as follows.
\begin%
{itemize}
\item
$\dif(\ctrue,\exp_1,\exp_2)$ is a pure redex whose reduct is $\exp_1$.
\item
$\dif(\cfalse,\exp_1,\exp_2)$ is a pure redex whose reduct is $\exp_2$.
\item
$\letin{\tuple{x_1,x_2}=\tuple{\val_1,\val_2}}{\exp}$
is a pure redex whose reduct is $\esubst{\val_1,\val_2}{x_1,x_2}{\exp}$.
\item
$\dfst{\tuple{\val_1,\val_2}}$ is a pure redex whose reduct is $\val_1$.
\item
$\dsnd{\tuple{\val_1,\val_2}}$ is a pure redex whose reduct is $\val_2$.
\item
$\app{\lam{x}{\exp}}{\val}$
is a pure redex whose reduct is $\esubst{\val}{x}{\exp}$.
\item
$\fix{x}{\val}$ is a pure redex whose reduct is $\esubst{\fix{x}{\val}}{x}{\val}$.
\end{itemize}
\end{definition}

\def\eval{\rightarrow}
\def\Eval{\Rightarrow}
\def\frandbit{\mbox{\it randbit}}
Evaluating calls to constant functions is of particular importance in
$\MTLCZ$. Assume that $\cfun$ is a constant function of arity $n$. The
expression $\cfun(v_1,\ldots,v_n)$ is an {\em ad-hoc} redex if $\cfun$ is
defined at $v_1,\ldots,v_n$, and any value of $\cfun(v_1,\ldots,v_n)$ is a
reduct of $\cfun(v_1,\ldots,v_n)$. For instance, $1+1$ is an ad hoc redex
and $2$ is its sole reduct. In contrast, $1+\ctrue$ is not a redex as it is
undefined. We can even have non-deterministic constant functions.  For
instance, we may assume that the ad-hoc redex $\frandbit()$ can evaluate to
both 0 and 1.

Let $\exp$ be a well-typed expression of the form
$\cfun(v_1,\ldots,v_n)$ and $\resof(\exp)\subseteq\rns$ holds for some
valid $\rns$ (that is, $\rns\in\RES$).  We always assume that there
exists a reduct $\val$ in $\MTLCZ$ for $\cfun(v_1,\ldots,v_n)$ such
that $(\rns\backslash\resof(e))\uplus\resof(v)\in\RES$. By doing so,
we are able to give a presentation with much less clutter.

\begin%
{definition} Given expressions $\exp_1$ and $\exp_2$, we write
$\exp_1\eval\exp_2$ if $\exp_1=E[\exp]$ and $\exp_2=E[\exp']$ for some
$E,\exp$ and $\exp'$ such that $\exp$ is a redex, $\exp'$ is a reduct
of $\exp$, and we may say that $\exp_1$ evaluates or reduces to
$\exp_2$ purely if $\exp$ is a pure redex.
\end{definition}

Note that resources may be generated as well as consumed when ad-hoc
reductions occur. This is an essential issue in any linear type system
designed to support practical programming.

\begin%
{definition}
Given two pools $\pool_1$ and $\pool_2$, the relation
$\pool_1\eval\pool_2$ is defined according to the following
rules:\\[-18pt]
\begin{center}
\[
\fontsize%
{8pt}{9pt}\selectfont
\begin{array}{c}
\infer[\mbox{(PR0)}]
      {\pool[\tid\mapsto\exp_1]\eval\pool[\tid\mapsto\exp_2]}
      {\exp_1\eval\exp_2}
\\[6pt]
\infer[\mbox{(PR1)}]
      {\pool\eval\pool[\tid_0:=E[\tuple{}]][\tid\mapsto\app{\lam{x}{e}}{\tuple{}}]}
      {\pool(\tid_0)=E[\fthreadcreate(\lam{x}{e})]}
\\[6pt]
\infer[\mbox{(PR2)}]
      {\pool[\tid\mapsto\tuple{}]\eval\pool}{\tid > 0} \\[2pt]
\end{array}\]
\end{center}
\end{definition}
If a pool $\pool_1$ evaluates to another pool $\pool_2$ by the rule (PR0),
then one thread in $\pool_1$ evaluates to its counterpart in $\pool_2$ and
the rest stay the same; if by the rule (PR1), then a fresh thread is
created; if by the rule (PR2), then a thread (that is not the main
thread) is terminated.

From this point on, we always (implicitly) assume that
$\resof(\pool)\in\RES$ holds whenever $\pool$ is well-typed.  The soundness
of the type system of $\MTLCZ$ rests upon the following two theorems:
\begin%
{theorem}
(Subject Reduction on Pools)
\label{theorem:MTLCZ:subject_reduction_on_pools}
Assume that $\tpjg\pool_1:\vwty$ is derivable and $\pool_1\eval\pool_2$
holds for some $\pool_2$ satisfying $\resof(\pool_2)\in\RES$. Then
$\tpjg\pool_2:\vwty$ is also derivable.
\end{theorem}
\begin%
{proof}
By structural induction on the derivation of $\tpjg\pool_1:\vwty$.
Note that Lemma~\ref{lemma:MTLCZ:substitution} is needed.
\end{proof}

\begin%
{theorem}
(Progress Property on Pools)
\label{theorem:MTLCZ:progress_on_pools}
Assume that $\tpjg\pool_1:\vwty$ is derivable. Then we have the following
possibilities:
\begin{itemize}
\item
$\pool_1$ is a singleton mapping $[0\mapsto\val]$ for some $\val$, or
\item
$\pool_1\eval\pool_2$ holds for some $\pool_2$ satisfying $\resof(\pool_2)\in\RES$.
\end{itemize}
\end{theorem}
\begin%
{proof}
By structural induction on the derivation of $\tpjg\pool_1:\vwty$.  Note
that Lemma~\ref{lemma:MTLCZ:canonical} is needed. Essentially, we can
readily prove that $\Pi_1(\tid)$ for any $\tid\in\dom(\Pi_1)$ is either a
value or of the form $E[\exp]$ for some evaluation context $E$ and redex
$\exp$.  If $\Pi_1(\tid)$ is a value for some $\tid>0$, then this value
must be $\tuple{}$ and the rule $\mbox{(PR2)}$ can be used to reduce
$\Pi_1$.  If $\Pi_1(\tid)$ is of the form $E[e]$ for some redex $e$, then
the rule $\mbox{(PR0)}$ can be used to reduce $\Pi_1$.
\end{proof}

By combining Theorem~\ref{theorem:MTLCZ:subject_reduction_on_pools}
and Theorem~\ref{theorem:MTLCZ:progress_on_pools}, we can immediately
conclude that the evaluation of a well-typed pool either leads to a
pool that is a singleton mapping of the form $[0\mapsto\val]$ for some
value $\val$, or it goes on forever.  In other words, $\MTLCZ$ is
type-sound.

\section%
{$\MTLCO$: Extending $\MTLCZ$ with multiparty channels}
\label{section:MTLCO}
\def\ST{\mbox{$S$}}
\def\stact{\mbox{\it act}}
\def\stnil{\mbox{\bf nil}}
\def\stmsg{\mbox{\bf msg}}
There is no support for communication between threads in $\MTLCZ$,
making $\MTLCZ$ uninteresting as a multi-threaded language. We extend
$\MTLCZ$ to $\MTLCO$ with support for synchronous communication
channels in this section. Supporting asynchronous communication
channels is certainly possible but would result in a more involved
theoretical development.  In order to assign types to channels, we
introduce session types as follows:
\[
\begin%
{array}{rcl}
\ST & ::= & \stmsg(r_0,r_1) \mid \ST_1\mconj_{r}\ST_2 \\
\end{array}
\]
where $r_0$ and $r_1$ range over distinct roles. For a simplified
presentation, we use $\stmsg(r_0,r_1)$ to mean messaging from role
$r_0$ to role $r_1$ (rather than $\stmsg(r_0,r_1,\vwty)$ for
specifying a value of viewtype $\vwty$ being sent from $r_0$ to
$r_1$). As for $\mconj_{r}$, its meaning is explained in
Section~\ref{subsection:mconj/mdisj}. For brevity, we skip session
types of the form $\ST_1\aconj_{r}\ST_2$.

Given a role set $R$ and a session type $\ST$, we can form a base
viewtype $\tchan(R,\ST)$ for an endpoint and refer to $R$ as the role
set attached to this endpoint. A process holding an endpoint of type
$\tchan(R,\ST)$ is supposed to implement all of the roles in $R$.

\def\fchancreate{\mbox{\tt chan\_create}}
The function $\fchancreate$ for creating a channel of two endpoints
is assigned the following c-type schema:
$$
\begin%
{array}{rcl}
\fchancreate & : & (\tchan(R, \ST) \ltimp \tunit) \ctimp \tchan(\setcomp{R}, \ST) \\
\end{array}
$$
In order to construct a channel of 3 or more endpoints, we can make
use of $\fchanthreecut$. Assume that each $\CH^i$ has two endpoints
$\CH^{i}_{R_i}$ and $\CH^{i}_{\setcomp{R}_i}$ for $i=1,2$.  If
$\setcomp{R}_1$ and $\setcomp{R}_2$ are disjoint, then calling
$\fchancreate$ on $\lam{x}{\fchanthreecut(\CH^1_{R_1},\CH^2_{R_2},x)}$
returns an endpoint $\CH^3_{R_1\cap R_2}$ such that the three
endpoints $\CH^1_{\setcomp{R_1}}$, $\CH^2_{\setcomp{R_2}}$, and
$\CH^3_{R_1\cap R_2}$ form a new channel.

Please recall that the function $\fchansync$ for $\stmsg$ is given
the following c-type schema:
$$
\begin%
{array}{rcl}
\fchansync & : & (\tchan(R, \tmsg(r_0,r_1))) \ctimp \tunit
\end{array}
$$
Given an endpoint $\CH_R$, $\fchansync$ sends a message at $\CH_R$
if $r_0\in R$ and $r_1\not\in R$, and it receives a message
if $r_0\not\in R$ and $r_1\in R$, and it does nothing otherwise.

There are no new typing rules in $\MTLCO$ over $\MTLCZ$.  In any type
derivation of $\pool:\vwty$, the types assigned to the endpoints of
any channel $\CH$ are required to match in the sense that there exists
$\ST$ such that these types are of form $\tchan(R_1, \ST),\ldots,
\tchan(R_n, \ST)$ for $R_1\uplus\ldots\uplus{R}_n=\fullset$. Clearly,
this can be seen as some kind of coherence requirement. For evaluating
pools in $\MTLCO$, we have the following additional rules:
$$
\fontsize%
{8pt}{9pt}\selectfont
\begin
{array}{c}
\infer%
[\hbox{(PR3)}]
{\pool\eval\mapadd{\maprep{\pool}{\tid_0}{E[\CH_{\setcomp{R}}]}}{\tid}{\esubst{x}{\CH_{R}}{e}}}
{\pool(\tid_0)=E[\fchancreate(\lam{x}{e})]}
\\[4pt]

\infer%
[\hbox{(PR4)}]
{\pool\eval\pool[\tid_1:=E_1[\dunit]]\ldots[\tid_n:=E_n[\dunit]]}
{R_1\uplus\ldots\uplus{R_n}=\fullset&\pool(\tid_1)=E_1[\fchansync(\CH_{R_1})]&\cdots&\pool(\tid_n)=E_n[\fchansync(\CH_{R_n})]}
\\[4pt]
\end{array}
$$
Due to the symmetry between $\fchanmdisjl$ and $\fchanmdisjr$, we only
consider the former in the following presentation. Another
rule for evaluating pools in $\MTLCO$ is given as follows:
$$
\fontsize%
{8pt}{9pt}\selectfont
\begin
{array}{c}
\infer%
[\hbox{(PR5)}]
{
\pool\eval
\pool_1%
[\tid_2:=E_2[\CH'_{R_2}]][\tid'_2\mapsto\esubst{\CH''_{R_2}}{x}{e_2}]%
\cdots%
[\tid_n:=E_n[\CH'_{R_n}]][\tid'_n\mapsto\esubst{\CH''_{R_n}}{x}{e_n}]%
}
{$$
\begin%
{array}{c}
R_1\uplus\ldots\uplus{R_n}=\fullset\kern10pt\pool(\tid_1)=E_1[\fchanmconj(\CH_{R_1})] \\
\pool(\tid_2)=E_2[\fchanmdisjl(\CH_{R_2},\lam{x}{e_2})]~\cdots~\pool(\tid_n)=E_n[\fchanmdisjl(\CH_{R_n},\lam{x}{e_n})] \\
\end{array}
$$}
\\[2pt]
\end{array}
$$
where
$\CH'$ and $\CH''$ are two new channels, and
$\pool_1=\pool[\tid_1:=E_1[\tuple{\CH'_{R_1},\CH''_{R_1}}]]$,
and $\tid'_2,\ldots,\tid'_n$ are newly created thread ids.

For brevity, we omit rules for other primitive functions on channels
(e.g., $\fchantwocut$ and $\fchanthreecut$), which can be formulated
by following $\mbox{(PR5)}$.

Theorem~\ref{theorem:MTLCZ:subject_reduction_on_pools} (Subject
Reduction) can be readily established for $\MTLCO$. As for
Theorem~\ref{theorem:MTLCZ:progress_on_pools}, we need some special
treatment due to the presence of session-typed primitive functions
such as $\fchancreate$, $\fchanmconj$, $\fchanmdisjl$, etc.

A partial (ad hoc) redex in $\MTLCO$ is an expression of one of the
following forms: $\fchansync(\CH_R)$, $\fchanmconj(\CH_R)$,
$\fchanmdisjl(\CH_R)$.  We can immediately prove in $\MTLCO$ that each
well-typed program is either a value or of the form $E[e]$ for some
evaluation context $E$ and expression $e$ such that $e$ is either a
redex or a partial redex. We refer to an expression as a blocked one
if it is of the form $E[e]$ for some partial redex $e$. A set of
partial redexes are said to be matching if
\begin%
{itemize}
\item
it consists of
$\fchansync(\CH_{R_1}),\ldots,\fchansync(\CH_{R_n})$
for $R_1\uplus\ldots\uplus{R}_n=\fullset$, or
\item
it consists of the following:
$$\fchanmconj(\CH_{R_1}),\fchanmdisjl(\CH_{R_2},v_2),\ldots,\fchanmdisjl(\CH_{R_n},v_n)$$
for $R_1\uplus\ldots\uplus{R}_n=\fullset$.
\end{itemize}
A set of blocked expressions are matching if their partial redexes
form a matching set.  Clearly, a pool containing a set of matching
blocked expressions can evaluate according to the rule $\mbox{(PR4)}$
or $\mbox{(PR5)}$. Therefore, a pool is deadlocked only when there is
no subset of matching expressions among the entire set of blocked
expressions.

\def\nchan{\mbox{\#chan}}
\def\nindep{\mbox{\#indep}}
\def\nendpt{\mbox{\#endpt}}
\def\sizegtz#1{|#1|^{+}}
\begin
{definition}
Assume there are $k$ channels $\CH^1,\ldots,\CH^k$ in a pool $\pool$
such that each channel $\CH^i$ consists of $n_i$ endpoints for
$i=1,\ldots,k$. Let us use $\nchan(\pool)$ for $k$ and
$\nendpt(\pool)$ for $\Sigma_{i=1}^{k} n_i$. In addition, let us use
$\sizegtz{\pool}$ for the number of threads in $\pool$ holding at
least one endpoint. We say that $\pool$ is \emph{relaxed} if
$\sizegtz{\pool}+\nchan(\pool)\geq\nendpt(\pool)+1$ holds.
\end{definition}

\begin
{lemma}
\label{lemma:relaxed:preserve}
If $\pool$ is obtained from evaluating an initial pool
containing no channels, then $\pool$ is relaxed.
\end{lemma}
\begin%
{proof}
By a careful inspection on the
evaluation rules $\mbox{(PR3)}$, $\mbox{(PR4)}$ and $\mbox{(PR5)}$.
\end{proof}

\begin%
{theorem}
(Progress Property on Pools)
\label{theorem:MTLCO:progress_on_pools}
Assume that $\tpjg\pool_1:\ty$ is derivable.
If $\pool_1$ is relaxed, then
either $\pool_1$ is a singleton mapping $[0\mapsto\val]$ for some
$\val$ or $\pool_1\eval\pool_2$ holds for some $\pool_2$.
\end{theorem}
\begin%
{proof}
Assume there are $k$ channels $\CH^1,\ldots,\CH^k$ in $\pool_1$ for
some $k>0$ such that each channel $\CH^i$ consists of $n_i$ endpoints
for $i=1,\ldots,k$.  Note that
$\sizegtz{\pool_1}+k\geq(\Sigma_{i=1}^{k} n_i)+1$ holds as $\pool_1$
is relaxed.  Assume that each thread in $\pool_1$ is a blocked
expression.  By the pigeonhole principle, there must be $n_i$ blocked
expressions involving $\CH^i$ for some $i$.  Since $\pool_1$ is
well-typed, these $n_i$ blocked expressions form a matching set,
allowing $\pool_1$ to evaluate to some $\pool_2$ according to the rule
$\mbox{(PR4)}$ or $\mbox{(PR5)}$.
\end{proof}
Note that Theorem~\ref{theorem:MTLCO:progress_on_pools} establishes a
form of global progress in the sense that multiple threads must
coordinate in order to make progress in evaluation.

Assume that we start with a well-typed pool $\pool$ containing no
channels.  By Lemma~\ref{lemma:relaxed:preserve}, any pool that is
reduced from $\pool$ is relaxed.  With subject reduction for $\MTLCO$
and Theorem~\ref{theorem:MTLCO:progress_on_pools}, we can conclude
that either $\pool$ evaluates to a singleton mapping $[0\mapsto\val]$
for some $\val$ or the evaluation goes on forever.

\def\fchantwocreate{\mbox{\tt chan2\_create}}
Please assume for the moment that we would like to add into $\MTLCO$
a function of the name $\fchantwocreate$ of the following type schema:
\[
\begin%
{array}{c}
((\tchan(R_1,\ST_1),\tchan(R_2,\ST_2))\ltimp\tunit)\ctimp(\tchan(\setcomp{R_1},\ST_1),\tchan(\setcomp{R_2},\ST_2)) \\
\end{array}\]
One may think of $\fchantwocreate$ as a ``reasonable'' generalization
of $\fchancreate$ that creates in a single call two endpoints instead
of one.  Unfortunately, adding $\fchantwocreate$ into $\MTLCO$ can
potentially cause a deadlock. For instance, a thread may wait for a
message on the first of the two endpoints $(\CH^1,\CH^2)$ returned
from a call to $\fchantwocreate$ while the newly created thread waits
for a message on the second argument of the function passed to create
it, resulting in a deadlocked situation where neither of the two
threads is able to send to the other one. Clearly, a pool cannot be
relaxed if it contains two threads and two channels such that each
channel consist of $2$ endpoints.

\section%
{Implementing Multiparty Channels}
\label{section:implementation}
We have implemented in ATS~\cite{ats-lang} the session-typed
multiparty channels presented in this paper.  As far as typechecking
is of the concern, the only considerable complication comes from the
need for solving constraints that involve various common set
operations (on role sets).  We currently export such constraints for
them to be solved with an external constraint-solver based on
Z3~\cite{Z3-tacas08}.

The first implementation is based on shared memory, which is used to
construct ATS programs that compile to C. Another implementation is
based on processes in Erlang, which is for use in ATS programs that
compile to Erlang (so as to take advantage of the infrastructural
support for distributed programming in Erlang).
For taking a peek at a running implementation of Example~1 in
Section~\ref{subsection:more_session_constructors}, please visit the
following link:
\begin%
{center}
\texttt{http://pastebin.com/JmZRukRi}
\end{center}
where the code should be accessible to someone familiar with ML-like
syntax. To facilitate understanding of this example, we outline as
follows some key steps involved in building a 3-party session that
makes genuine use of non-singleton role sets.

\subsection%
{A Sketch for Building a 3-Party Session}
\label{subsection:sketch:3-party_session}
\def\tservice{{\bf service}}
\def\fservicecreate{\mbox{\tt service\_create}}
\def\fservicerequest{\mbox{\tt service\_request}}
Building a session often requires explicit coordination between the
involved parties during the phase of setting-up.  In practice,
designing and implementing coordination between 3 or more parties is
generally considered a difficult issue. By building a multiparty
channel based on 2-party channels, we only need to be concerned with
two-party coordination, which is usually much easier to handle.

Given a role set $R$ and a session type $\ST$, we introduce a type
$\tservice(R,\ST)$ that can be assigned to a value representing some
form of {\it persistent} service.  With such a service, channels of
type $\tchan(R,\ST)$ can be created repeatedly. A built-in function
$\fservicecreate$ is assigned the following type for creating a
service:
$$
(\tchan(\setcomp{R},\ST)\itimp\tunit)\ctimp\tservice(R,\ST)
$$
In contrast with $\fchancreate$ for creating a channel, $\fservicecreate$
requires that its argument be a non-linear function (so that this function
can be called repeatedly).
A client may call the following function to obtain a channel
to communicate with a server that provides the requested service:
\[\begin%
{array}{rcl}
\fservicerequest & : & (\tservice(R,\ST))\ctimp\tchan(R,\ST) \\
\end{array}\]
Suppose we want to build a 3-party session involving 3 roles: 0, 1,
and 2.  We may assume that there are two services of types
$\tservice(\myset{1,2},\ST)$ and $\tservice(\myset{0,2},\ST)$
available to a party (planning to implement role 2); this party can
call $\fservicerequest$ on the two services (which are just two names)
to obtain two channels $\CH^0$ and $\CH^1$ of types
$\tchan(\myset{1,2},\ST)$ and $\tchan(\myset{0,2},\ST)$, respectively;
it then calls $\fchantwocutres(\CH^0,\CH^1)$ to obtain a channel
$\CH^2$ of type $\tchan(\myset{2},\ST)$ for communicating with two
servers providing the requested services. There are certainly many
other ways of building a multiparty channel by passing around 2-party
channels.

\section%
{Related Work and Conclusion}
Multirole as a logical notion generalizes the notion of duality in
logic. One may compare multirole to the notion of linearity in linear
logic~\cite{LinearLogic} as multirole/linearity can often be
incorporated into an existing logic to form a multirole/linear version
of the logic. In this sense, multirole may be referred to as a logical
aspect. For instance, it should not be surprising if modal operators
can be generalized to ultrafilters as well even though we have not yet
formulated any multirole modal logic.

The admissibility of (binary) cut-rule in classical logic and
intuitionistic logic goes back to Gentzen's seminal work on LK and
LJ~\cite{SequentCalculus}.  We hereby generalize cut-rule to 1-cut and
2-cut-with-residual, which in turn yield a cut-rule (mp-cut) involving
$n$ sequents for any $n\geq 1$. The proof we give for the
admissibility of 2-cut-with-residual is directly based on one for the
admissibility of cut-rule in classical linear
logic~\cite{LL-Troelstra92}.

Session types were introduced by Honda~\cite{Honda93} and further
extended subsequently~\cite{TakeuchiHK94,HondaVK98}.  There have since
been extensive theoretical studies on session types in the
literature(e.g.,~\cite{CastagnaDGP09,GayVascon10,ToninhoCP11,Vasconcelos12,LindleyM15}).
Multiparty session types, as a generalization of (dyadic) session
types, were introduced by Honda and others~\cite{HondaYC08}, together
with the notion of global types, local types, projection and
coherence.

The session types presented in Section~\ref{section:MTLCO} are
global. In the future, we plan to introduce projection of global
session types into local ones. For instance, the local session type
for Contestant1 in Example~3 should be given as follows:
\begin%
{center}
\begin%
{minipage}{9cm}
\begin%
{verbatim}
query(0)@(mconj(0, answer(1,0)@score(0,1), nil))
\end{verbatim}
\end{minipage}\end{center}\vspace{2pt}
as there is no need for Contestant1(1) to learn the protocol that
solely specifies communication between Judge(0) and Contestant2(2).
Naturally, a set of local session types are coherent if they can be
projected from a global session type with respect to some role sets
$R_1,\ldots,R_n$ satisfying $R_1\uplus\ldots\uplus R_n=\fullset$.

In~\cite{Carbone:2017ki}, a Curry-Howard correspondence is proposed
between a language for programming multiparty sessions and a
generalization of CLL, where propositions correspond to the local
behavior of a party, proofs to processes, and proof normalization to
executing communications. In particular, Multiparty Classical
Processes (MCP) is developed that generalizes the duality in CLL to a new
notion of n-ary compatibility (referred to as coherence) and the cut
rule of CLL to a new rule for composing processes. Compared with MCP,
we see that our work on multirole logic presents a fundamentally
different approach to generalizing the notion of duality in logic (not
just in CLL). For instance, we give an interpretation of $A\otimes B$
based on LMRL as interleaving of $A$ and $B$ in arbitrary order and
$A\invamp B$ as behaving like $A$ and $B$ concurrently while MCP
interprets $\otimes$ and $\invamp$ as input and output, respectively.
There is little in common between LMRL and MCP except for both sharing
a similar motivation on generalizing duality in logic.

There is also very recent work on encoding multiparty session types
based on binary session types~\cite{CairesPerez16}, which relies on an
arbiter process to mediate communications between multiple parties
while preserving global sequencing information.  We see that this form
of mediating (formulated based on automata theory) is closely related
to performing a multiparty cut.

\def\SILL{\mbox{SILL}}
Probably $\MTLCO$ is most closely related to
$\SILL$~\cite{ToninhoCP13}, a functional programming language that
adopts via a contextual monad a computational interpretation of
intuitionistic linear sequent calculus as session-typed processes. The
approach we use to establish global progress for $\MTLCO$ can be seen
as an adaptation of one used in $\SILL$ for the same purpose. Unlike
$\MTLCO$, there are only binary channels in $\SILL$ and the support of
linear types in $\SILL$ is not direct and only monadic values
(representing open process expressions) can be linear.

Also, $\MTLCO$ is related to previous work on incorporating session
types into a multi-threaded functional
language~\cite{VasconcelosRG04}, where a type safety theorem is
established to ensure that the evaluation of a well-typed program can
never lead to a so-called {\em faulty configuration}. However, this
theorem does not imply global progress as a program that is not of
faulty configuration can still deadlock.  Also, we point out that
$\MTLCO$ is related to recent work on assigning an operational
semantics to a variant of GV~\cite{LindleyM15}, which takes a similar
approach to establishing deadlock-freedom (that is, global progress).

Given $\MRLJ$ and $\LMRL$,
it seems straightforward to formulate
$\LMRLJ$ as a multirole version of intuitionistic linear logic.
As session types can also be directly based on intuitionistic linear
propositions~\cite{CairesPfenning10}, it is only natural to also study
multiparty channels in $\LMRLJ$.

There are a variety of programming issues that need to be addressed in
order to facilitate the use of session types in practice. Currently,
session types are often represented as datatypes in ATS, and
programming with such session types tends to involve writing a very
significant amount of boilerplate code (as can be seen in the lengthy
implementation of the S0B1B2 example). In the presence of large and
complex session types, writing such code can be tedious and
error-prone. Naturally, we are interested in developing some
meta-programming support for generating such code automatically.
Also, we are in the process of designing and implementing session
combinators (in a spirit similar to parsing
combinators~\cite{Hutton92}) that can be conveniently called to
assemble subsessions into a coherent whole.






\appendix
\section{Appendix}

\subsection
{
Proof of Lemma~\ref{lemma:LMRL:2-cut-residual}
}
\label{Appendix:subsection:LMRL:2-cut-residual}

Assume $\D_1::(\Gamma_1,\formset{\interp{R_1}{A}})$ and
$\D_2::(\Gamma_2,\formset{\interp{R_2}{A}})$.  We proceed by induction
on $\sizeof{A}$ (the size of $A$) and $\height{\D_1}+\height{\D_2}$,
lexicographically ordered. For brevity, we are to focus only on the
most interesting case where there is one occurrence of
$\interp{R_i}{A}$ in $\formset{\interp{R_i}{A}}$ that is the major
formula of the last rule applied in $\D_i$, where $i$ ranges over $1$
and $2$. For this case, we have several subcases covering all the
possible forms that $A$ may take.

\subsubsection
{Assume that $A$ is a primitive formula}
It is a simple (but very meaningful) routine to verify that the
sequent $\tpjg(\Gamma_1,\Gamma_2,\interp{R_1\cap R_2}{A})$ follows
from an application of the rule $\mbox{\bf(Id)}$.

\subsubsection
{Assume that $A$ is pf the form $\lneg_{f}(B)$}
Then $\D_k$ is of the following form
for each of the cases $k=1$ and $k=2$:
$$
\infer[\mbox{\bf($\lneg$)}]
{\tpjg\Gamma_k,\interp{R_k}{A}}
{\D_{1k}::(\Gamma_{1k},\interp{\preimg{f}{R_k}}{B})}
$$  
By induction hypothesis on $\D_{11}$ and $\D_{12}$,
we obtain a derivation:
$$\D'::\Gamma_1,\Gamma_2,\interp{\preimg{f}{R_1}\cap\preimg{f}{R_2}}{B}$$
Note that the
$\preimg{f}{R_1\cap{R}_2}=\preimg{f}{R_1}\cap\preimg{f}{R_2}$.
By applying the rule $\mbox{\bf($\lneg$)}$ to $\D'$, we derive
$\tpjg\Gamma,\interp{R_1\cap R_2}{A}$.

\subsubsection
{Assume that $A$ is of the form $\lmrlmc{\U}{A_1}{A_2}$}
We have three possibilities: $R_1\in\U$ and $R_2\not\in\U$, or
$R_1\not\in\U$ and $R_2\in\U$, or $R_1\in\U$ and $R_2\in\U$.
\begin%
{itemize}
\item
Assume $R_1\in\U$ and $R_2\not\in\U$.
Then $\D_1$ is of the following form:
$$
\infer[\mbox{\bf($\mconj$-pos)}]
{\tpjg\Gamma_1,\interp{R_1}{A}}
{\D_{11}::(\Gamma_{11},\interp{R_1}{A_1})
 \kern12pt
 \D_{12}::(\Gamma_{12},\interp{R_1}{A_2})}
$$
and $\D_2$ is of the following form:
$$
\infer[\mbox{\bf($\mconj$-neg)}]
{\tpjg\Gamma_2,\interp{R_2}{A}}
{\D_{21}::(\Gamma_2,\interp{R_2}{A_1},\interp{R_2}{A_2})}
$$
By the induction hypothesis on $\D_{11}$ and $\D_{21}$, we have
a derivation:
$$
\D'_{11}::(\Gamma_{11},\Gamma_2,\interp{R_1\cap R_2}{A_1},\interp{R_2}{A_2})
$$
By the induction hypothesis on $\D_{12}$ and $\D'_{11}$, we have
a derivation:
$$
\D'_{12}::(\Gamma,\interp{R_1\cap R_2}{A_1},\interp{R_1\cap R_2}{A_2})
$$
By applying the rule $\mbox{\bf($\mconj$-neg)}$ to $\D'_{12}$, we
derive $\tpjg\Gamma,\interp{R_1\cap R_2}{A}$.
\item
Assume $R_1\not\in\U$ and $R_2\in\U$.
Then this case is analogous to the previous one.
\item
Assume $R_1\in\U$ and $R_2\in\U$.
Then $\D_k$ is of the following form
for each of the cases $k=1$ and $k=2$:
$$
\infer[\mbox{\bf($\mconj$-pos)}]
{\tpjg\Gamma_k,\interp{R_k}{A}}
{\D_{k1}::(\Gamma_{k1},\interp{R_k}{A_1})
 \kern12pt
 \D_{k2}::(\Gamma_{k2},\interp{R_k}{A_2})}
$$
By the induction hypothesis on $\D_{11}$ and $\D_{21}$,
we obtain a derivation:
$$\D'_{1}::(\Gamma_{11},\Gamma_{21},\interp{R_1\cap R_2}{A_1})$$
By the induction hypothesis on $\D_{12}$ and $\D_{22}$,
we obtain a derivation:
$$\D'_{2}::(\Gamma_{12},\Gamma_{22},\interp{R_1\cap R_2}{A_2})$$
By applying the rule $\mbox{\bf($\mconj$-pos)}$ to $\D'_{1}$ and
$\D'_{2}$, we derive $\tpjg\Gamma,\interp{R_1\cap R_2}{A}$.
\end{itemize}

\subsubsection
{Assume that $A$ is of the form $\lmrlac{r}{A_1}{A_2}$}
We have three possibilities: $R_1\in\U$ and $R_2\not\in\U$, or
$R_1\not\in\U$ and $R_2\in\U$, or $R_1\in\U$ and $R_2\in\U$.
\begin%
{itemize}
\item
Assume $R_1\in\U$ and $R_2\not\in\U$.
Then $\D_1$ is of the following form:
$$
\infer[\mbox{\bf($\&$-pos)}]
{\tpjg\Gamma_1,\interp{R_1}{A}}
{\D_{11}::(\Gamma_1,\interp{R_1}{A_1})
 \kern12pt
 \D_{12}::(\Gamma_1,\interp{R_1}{A_2})}
$$
and $\D_2$ is of the following form for $k$ being either $1$ or $2$:
$$
\infer[]
{\tpjg\Gamma_2,\interp{R_2}{A}}
{\D_{2k}::(\Gamma_2,\interp{R_2}{A_k})}
$$
where the last applied rule in $\D_2$ is
\mbox{\bf($\&$-neg-l)} or \mbox{\bf($\&$-neg-r)}.
By induction hypothesis on $\D_{1k}$ and $\D_{2k}$,
we obtain a derivation:
$$\D'_{k}::(\Gamma_1,\Gamma_2,\interp{R_1\cap R_2}{A_k})$$
By applying to $\D'_{k}$
either $\mbox{\bf($\&$-neg-l)}$ or $\mbox{\bf($\&$-neg-r)}$,
we derive $\tpjg\Gamma_1,\Gamma_2,\interp{R_1\cap R_2}{A}$.
\item
Assume $R_1\not\in\U$ and $R_2\in\U$.
Then this case is analogous to the previous one.
\item
Assume $R_1\in\U$ and $R_2\in\U$.
Then $\D_k$ is of the following form
for each of the cases $k=1$ and $k=2$:
$$
\infer[\mbox{\bf($\&$-pos)}]
{\tpjg\Gamma_k,\interp{R_k}{A}}
{\D_{k1}::(\Gamma_{k1},\interp{R_k}{A_1})
 \kern12pt
 \D_{k2}::(\Gamma_{k2},\interp{R_k}{A_2})}
$$
By the induction hypothesis on $\D_{11}$ and $\D_{21}$,
we obtain a derivation:
$$\D'_{1}::(\Gamma_{11},\Gamma_{21},\interp{R_1\cap R_2}{A_1})$$
By the induction hypothesis on $\D_{12}$ and $\D_{22}$,
we obtain a derivation:
$$\D'_{2}::(\Gamma_{12},\Gamma_{22},\interp{R_1\cap R_2}{A_2})$$
By applying the rule $\mbox{\bf($\&$-pos)}$ to $\D'_{1}$ and $\D'_{2}$,
we derive
$\tpjg\Gamma_{1},\Gamma_{2},\interp{R_1\cap R_2}{A}$.
\end{itemize}

\subsubsection
{Assume that $A$ is of the form $\lmrlx{\U}{B}$}
This is the most involved subcase.
We have three possibilities: $R_1\in\U$ and $R_2\not\in\U$, or
$R_1\not\in\U$ and $R_2\in\U$, or $R_1\in\U$ and $R_2\in\U$.
\begin%
{itemize}
\item
Assume $R_1\in\U$ and $R_2\not\in\U$.
Then $\D_1$ is of the following form:
$$
\infer%
[\mbox{\bf($\bang$-pos)}]
{\tpjg\qmark(\Gamma_1),\interp{R_1}{A}}
{\D_{11}::\qmark(\Gamma_1),\interp{R_1}{B}}
$$
There are the following three possibilities for $\D_2$:
\begin%
{itemize}
\item
$\D_2$ is of the following form:
$$
\infer%
[\hbox{\mbox{\bf($\bang$-neg-weaken)}\hss}]
{\tpjg\Gamma_2,\formset{\interp{R_2}{A}}}
{\D_{21}::\Gamma_2,\formset{\interp{R_2}{A}}}
$$
We obtain a derivation of $(\Gamma_1,\Gamma_2,\interp{R_1\cap R_2}{A})$
by the induction hypothesis on $\D_1$ and $\D_{21}$.
\item
$\D_2$ is of the following form:
$$
\infer%
[\hbox{\mbox{\bf($\bang$-neg-derelict)}\hss}]
{\tpjg\Gamma_2,\formset{\interp{R_2}{A}}}
{\D_{21}::\Gamma_2,\formset{\interp{R_2}{A}},\interp{R}{B}}
$$
By the induction hypothesis on $\D_1$ and $\D_{21}$, we obtain
a derivation:
$$\D_{121}::(\Gamma_1,\Gamma_2,\interp{R_1\cap R_2}{A},\interp{R}{B})$$
By the induction hypothesis on $\D_{11}$ and $\D_{121}$, we obtain a
derivation:
$$\D'_{121}::(\Gamma_1,\Gamma_1,\Gamma_2,\interp{R_1\cap R_2}{A})$$
By applying the rule $\mbox{\bf($\bang$-neg-contract)}$ to $\D'_{121}$
repeatedly, we derive
$\tpjg\Gamma_1,\Gamma_2,\interp{R_1\cap R_2}{A}$.
\item
$\D_2$ is of the following form:
$$
\infer%
[\hbox{\mbox{\bf($\bang$-neg-contract)}\hss}]
{\tpjg\Gamma_2,\formset{\interp{R_2}{A}}}
{\D_{21}::\Gamma_2,\formset{\interp{R_2}{A}},\interp{R_2}{A}}
$$
We obtain a derivation of
$(\Gamma_1,\Gamma_2,\interp{R_1\cap R_2}{A})$ by the induction
hypothesis on $\D_1$ and $\D_{21}$.
\end{itemize}
\item
Assume $R_1\not\in\U$ and $R_2\in\U$.
This subcase is completely analogous to the previous one.
\item
Assume $R_1\in\U$ and $R_2\in\U$.
Then
$\D_k$ is of the following form for
each of the cases $k=1$ and $k=2$:
$$
\infer%
[\mbox{\bf($\bang$-pos)}]
{\tpjg\qmark(\Gamma_k),\interp{R_k}{A}}
{\D_{k1}::\qmark(\Gamma_k),\interp{R_k}{B}}
$$
We obtain
$\D'_{12}::(\qmark(\Gamma_1),\qmark(\Gamma_2),\interp{R_1\cap R_2}{B})$
by the induction hypothesis on $\D_{11}$ and $\D_{21}$.
We then obtain a derivation
of $(\qmark(\Gamma_1),\qmark(\Gamma_2),\interp{R_1\cap R_2}{A})$
by applying the rule $\mbox{\bf($\bang$-pos)}$ to $\D'_{12}$.
\end{itemize}

\subsubsection
{Assume that $A$ is of the form $\forall_{\U}(\lambda x.B)$}
We have three possibilities: $R_1\in\U$ and $R_2\not\in\U$,
or $R_1\not\in\U$ and $R_2\in\U$, or $R_1\in\U$ and $R_2\in\U$.
\begin%
{itemize}
\item
Assume $R_1\in\U$ and $R_2\not\in\U$.
Then $\D_1$ is of the following form:
$$
\infer%
[\mbox{\bf($\forall$-pos)}]
{\tpjg\Gamma,\interp{R_1}{A}}
{\D_{11}::(\Gamma_1,\interp{R_1}{B})}
$$
where $x$ does not have any free occurrences in $\Gamma_1$, and
$\D_2$ is of the following form:
$$
\infer
[\mbox{\bf($\forall$-neg)}]
{\tpjg\Gamma,\interp{R_2}{A}}
{\D_{21}::(\Gamma_2,\interp{R_2}{\subst{t}{x}{B}})}
$$
Let $\D'_{11}$ be $\subst{t}{x}{\D_{11}}$, which is a derivation of
$(\Gamma_1,\interp{R_1}{\subst{t}{x}{B}})$ obtained from substituting
$t$ for every free occurrence of $x$ in $\D_{11}$.  Clearly,
$\D'_{11}$ is of the same height as $\D_{11}$.  By the induction
hypothesis on $\D'_{11}$ and $\D_{21}$, we have a derivation:
$$\D_{121}::(\Gamma_1,\Gamma_2,\interp{R_1\cap R_2}{\subst{t}{x}{B}})$$
By applying the rule $\mbox{\bf($\forall$-neg)}$ to $\D_{121}$, we have
a derivation of $(\Gamma_1,\Gamma_2,\interp{R_1\cap R_2}{A})$.
\item
Assume $R_1\not\in\U$ and $R_2\in\U$.
Then this case is analogous to the previous one.
\item
Assume $R_1\in\U$ and $R_2\in\U$. 
Then $\D_k$ is of the following form
for each of the cases $k=1$ and $k=2$:
$$
\infer%
[\mbox{\bf($\forall$-pos)}]
{\tpjg\Gamma_k,\interp{R_k}{A}}
{\D_{k1}::(\Gamma_k,\interp{R_k}{A},\interp{R_k}{B})}
$$
where $x$ does not have free occurrences in $\Gamma_k$.
By the induction hypothesis on $\D_{11}$ and $\D_{21}$, we have
a derivation:
$$\D'_{12}::(\Gamma_1,\Gamma_2,\interp{R_1\cap R_2}{B})$$
By applying the rule $\mbox{\bf($\forall$-pos)}$ to $\D'_{12}$,
we obtain a derivation of $(\Gamma_1,\Gamma_2,\interp{R_1\cap R_2}{A})$.
\end{itemize}
All of the cases are covered where the cut-formula is the major formula
of both $\D_1$ and $\D_2$. For brevity, we omit the
cases where the cut-formula is not
the major formula of either $\D_1$ or $\D_2$, which can be trivially
handled~\cite{LL-Troelstra92}.

\bibliographystyle{jfp}\bibliography{mysub}
\end{document}